\documentclass[traditabstract]{aa}
\usepackage{graphicx}
\usepackage{natbib}
\usepackage{txfonts}

\begin{document}

\title{Evidence for a proto-black hole and a double astrophysical component in GRB 101023}
\titlerunning{A double astrophysical component in GRB 101023}
\authorrunning{A. V. Penacchioni et al.}
\author{A.V. Penacchioni\inst{1,3}, R. Ruffini\inst{1,2,3}, L. Izzo\inst{1}, M. Muccino\inst{1}, C.L. Bianco\inst{1,2}, L. Caito\inst{1,2}, B. Patricelli\inst{1}, L. Amati\inst{4,2}}
\institute{Dip. di Fisica, Sapienza Universit\`a di Roma and ICRA, Piazzale
Aldo Moro 5, I-00185 Rome, Italy. E-mail: [ana.penacchioni;ruffini;luca.izzo;marco.muccino;bianco;letizia.caito]@icra.it
\and
ICRANet, Piazzale della Repubblica 10, I-65122 Pescara, Italy. E-mail: barbara.patricelli@icranet.org
\and
Universit\'e de Nice Sophia Antipolis, Nice, CEDEX 2, Grand Chateau Parc Valrose,
\and
Italian National Institute for Astrophysics (INAF) - IASF Bologna, via P. Gobetti 101, 40129 Bologna, Italy. E-mail: amati@iasfbo.inaf.it}

\abstract{
\emph{Context:} It has been recently shown that GRB 090618, observed by AGILE, Coronas Photon, Fermi, Konus, Suzaku and Swift, is composed of two very different components: episode 1, lasting 50 s, shows a thermal plus power-law spectrum with a characteristic temperature evolving in time as a power law; episode 2 (the remaining 100 s) is a canonical long GRB. We have associated episode 1 to the progenitor of a collapsing bare core leading to the formation of a black hole: what was defined as a ``proto black hole''.\\
\emph{Aims:} In precise analogy with GRB 090618 we aim to analyze the 89s of the emission of GRB 101023, observed by Fermi, Gemini, Konus and Swift, to see if there are two different episodes: the first one presenting a characteristic black-body temperature evolving in time as a broken power law, and the second one consistent with a canonical GRB.\\
\emph{Methods:} To obtain information on the spectra, we analyzed the data provided by the GBM detector onboard the Fermi satellite, and we used the heasoft package XSPEC and RMFIT to obtain their spectral distribution. We also used the numerical code GRBsim to simulate the emission in the context of the fireshell scenario for episode 2.\\
\emph{Results:} We confirm that the first episode can be well fit by a black body plus power-law spectral model. The temperature changes with time following a broken power law, and the photon index of the power-law component presents a soft-to-hard evolution. We estimate that the radius of this source increases with time with a velocity of $1.5 \times 10^4 km/s$. The second episode appears to be a canonical GRB. By using the Amati and the Atteia relations, we determined the cosmological redshift, $z \sim 0.9 \pm 0.084 (stat.) \pm 0.2 (sys.)$. The results of GRB 090618 are compared and contrasted with the results of GRB 101023. Particularly striking is the scaling law of the soft X-ray component of the afterglow.\\
\emph{Conclusions:} We identify GRB 090618 and GRB 101023 with a new family of GRBs related to a single core collapse and presenting two astrophysical components: a first one related to the proto-black hole prior to the process of gravitational collapse (episode 1), and a second one, which is the canonical GRB (episode 2) emitted during the formation of the black hole. For the first time we are witnessing the process of a black hole formation from the instants preceding the gravitational collapse up to the GRB emission. This analysis indicates progress towards developing a GRB distance indicator based on understanding the P-GRB and the prompt emission, as well as the soft X-ray behavior of the late afterglow.
}


\date{}

\maketitle

\section{Introduction}\label{sec:1}

Discovered at the end of the 60s \citep{Strong1975}, gamma-ray bursts (GRBs) are extremely intense flashes of hard X-radiation, coming from random directions in the sky at unpredictable times and typically lasting from a fraction of a second up to a few minutes. They are detected by satellites in low Earth orbit at a rate of $\sim$0.8 events/day. As outlined by breakthrough observations in the last $\sim$15 years, these phenomena are by far the most energetic sources in the Universe, observed in a range of cosmological redshift $0.0084 \leq z \lesssim 9$ \citep{Salvaterra,Tanvir,Cucchiara}, with isotropic equivalent radiated energy $E_{iso}$ in the range $10^{49} - 10^{55}$ erg and a theoretically predicted upper limit to their energies of $10^{55}$ erg \citep{RuffiniMG12}. Since the early observation by BATSE \citep{Meegan1}, they have been divided into two classes: the short GRBs, with a characteristic duration of $T_{90} < 2$ s, and the long GRBs, with a characteristic $T_{90} > 2$ s \citep{Dezalay,Klebesadel,Kouveliotou}.
 
Analysis of the GRBs within the fireshell model \citep[see e.g.][and references therein]{Ruffini2001,Ruffini2009} has led to identifying a canonical GRB structure described by two parameters: the total energy $E_{tot}^{e^{\pm}}$ of the initially optically thick electron-positron plasma and its baryon load $B=M_B c^2/E_{tot}^{e^{\pm}}$. To this information characterizing the source is added the information on the density and filamentary distribution of the circumburst medium (CBM) \citep{Ruffini2004b,Ruffini2005,Patricelli2010,Patricelli2011}.

Within this model the structure of a canonical GRB has been identified. It is composed by a proper-GRB (P-GRB), followed by an extended afterglow. The P-GRB originates at the moment of transparency of the relativistically expanding electron-positron plasma.  The extended afterglow originates in the collision of the ultra-relativistic baryons with the filamentary structure of the CBM. The acceleration process of the baryons occurs in the optically thick phase of the self-accelerating electron-positron plasma. This explains the spiky emission observed in the prompt radiation \citep{Ruffini2002}. The average density, the porosity, and the dimensions of the clouds in the CBM are in turn determined \citep[see e.g.][]{Ruffini2006,Bernardini,Dainotti2007,Caito2009}.

This model has allowed the nature of long GRBs to be explained and two new classes of short bursts to be introduced. A first class contains the disguised short GRBs \citep{Bernardini,Caito2009,Caito2010,DeBarros}: just long GRBs exploding in low density CBM ($n=10^{-3} part/cm^3$), and often referred to as short GRBs in the literature \citep[see e.g.][]{Gehrels2005}. A second class contains the genuine short GRBs, theoretically foreseen in \citet{Ruffini2001} as canonical GRBs occurring in the limit of a very low baryon load, $B<10^{-5}$. This new class of genuine short GRBs is expected to occur on a much shorter time scale, $T_{90} \leq 10^{-2} - 10^{-3}$ s.
 
With the observation of GRB 090618, a novel situation has occurred with respect to the above classification. It had been shown in the pioneering works of Felix Ryde and his collaborators \citep{Ryde2004} that, in the early emission of selected BATSE sources and also in some Fermi sources, a characteristic thermal component is present with temperature changing in time following a broken power law  \citep{Ryde2004,Ryde2005,RydePeer2009}. They attempted to interpret this emission within the GRB fireball model \citep[see e.g.][]{Pe'er2007}.

\citet{Ruffini2010a} showed that two very different episodes occur in GRB 090618: episodes 1 and 2. Episode 1 presents an emission ``\'a la Ryde''. There it was proposed that such an emission, alternatively to the Ryde interpretation, had to be interpreted as originating in a new kind of source in the late phase of a core collapse. The concept of proto-black hole was introduced there. Episode 2 was shown to be consistent with a canonical GRB.

Details of the data analysis showing the characteristic broken power law temporal variation of the temperature of the thermal component of episode 1 are presented in \citet{Luca}. The radius of the emitting region and its time variation have been determined as well, along with the details of the GRB emission of episode 2, including the P-GRB structure, the porosity of the interstellar medium, the baryon load $B$, and the total energy. Identifying these two components has been made possible by the extraordinary coincidence of three major factors for this GRB: 1) precise determination of the cosmological redshift of this source $z=0.54$, implying the fortunate occurrence of a very close source with an energy $E_{iso} = 2.7 \times 10^{53}$ erg;
2) joint observations by several X and gamma-ray telescopes;
3) the exceptional dataset on the instantaneous spectral distribution, light curve, and luminosity variation of this source (see section \ref{GRB090618}).

\begin{figure}
\centering
\includegraphics[width=\hsize]{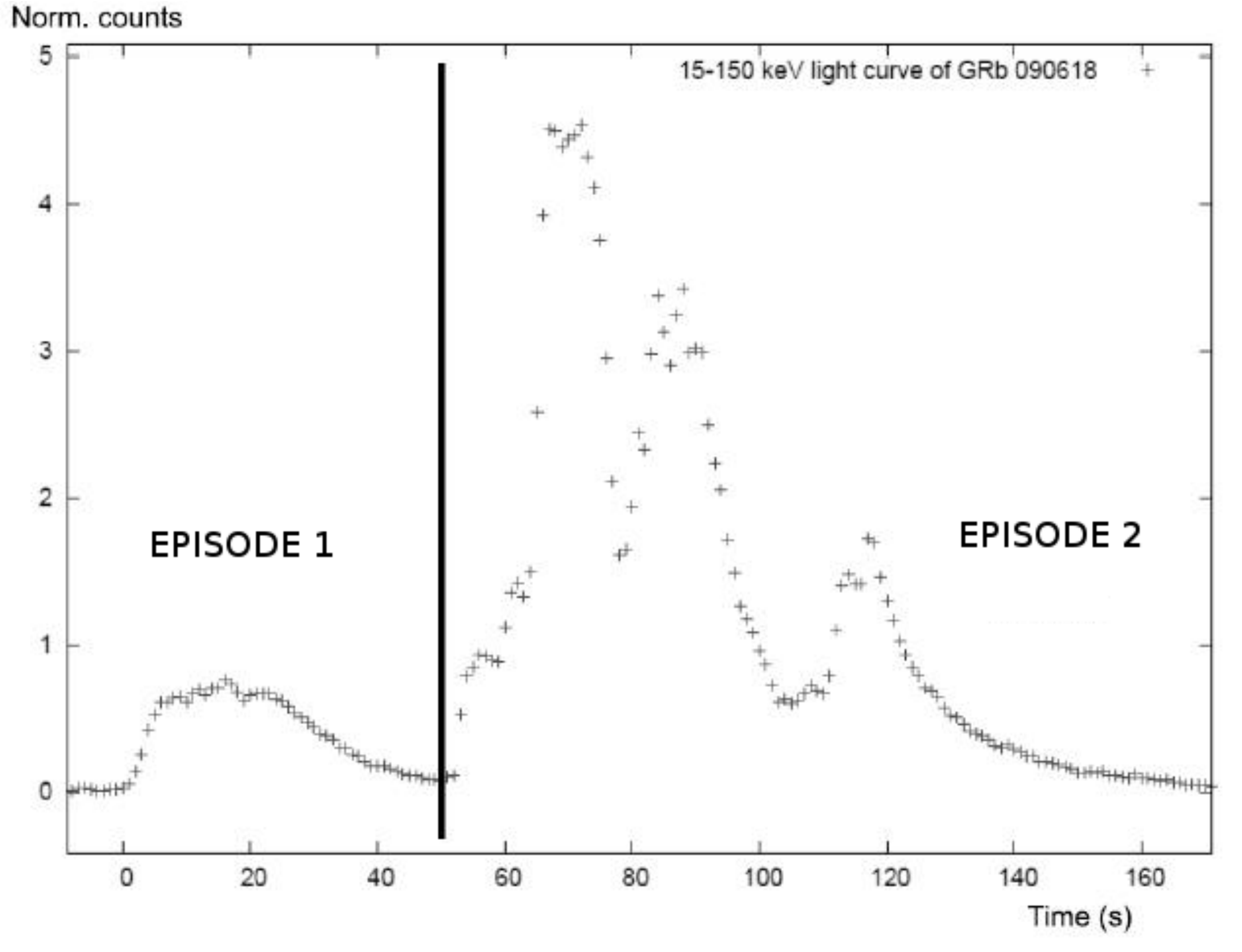}
\caption{Count light curve of GRB 090618 obtained from Fermi GBM detector, with a bin time of 1 s, and showing two-episode nature of the GRB.}
\label{lc090618}
\end{figure}

There is a striking morphological analogy between GRB 101023 and GRB 090618 (see Figs. \ref{picture of the light curve} and \ref{lc090618}). Both light curves present a first emission that lasts $\sim50$ s, followed by a spikier structure in the remaining part. We identify the first 45 s of GRB 101023 with episode 1 and the remaining 44 s with episode 2 (a canonical GRB). There is, however, a substantial difference between these two sources. In the present source, GRB 101023, the cosmological redshift is unknown. This has not been a drawback for us but a challenge that probes our understanding of the GRB phenomenon. It is interesting, as a rough estimate, that if one were to assume that the two sources, GRB 101023 and GRB 090618 had not only the same morphology but also the same energy $E_{iso}$, one would infer $z=1$ for the cosmological redshift of GRB 101023. A main result of this article is that, assuming the validity of the Amati relation \citep[see][and references therein]{Amati(2009)} and Atteia criteria \citep{Atteia}, it is possible to theoretically derive an expected cosmological redshift $z=0.9 \pm 0.084(stat.) \pm 0.2 (sys.)$ for episode 2. 

What is most striking is that we can have an independent verification of this redshift by comparing the late part of the afterglows of the two sources. Since we have verified that both GRB 090618 and GRB 101023 have similar energetics, and under the hypothesis of the same progenitor mechanism, we compare and contrast the luminosities of both GRBs in the late X-ray afterglow emission. 
We know that the X-ray afterglow is related to the residual kinetic energy of the outflow, although we do not attempt here to present a theoretical model for this emission.
We rescaled, in the observed time interval and energy range, the X-ray afterglow luminosity of GRB 090618 for different redshifts in an interval between $0.04 < z < 3$ (see Fig. \ref{fig:no29}). The striking coincidence for $z=0.9$ is presented in Fig. \ref{fig:no28}. 

In section \ref{GRB090618} we summarize the results of GRB 090618 and identify episode 1 and episode 2. 
In section \ref{sec:2} we present the observations of GRB 101023 by the different satellites.
In section \ref{sec:4} we give a brief summary of the fireshell scenario.
In section \ref{sec:5} we perform a spectral analysis of episodes 1 and 2 of this GRB. 
In section \ref{sec:6} we try to identify the P-GRB of the gamma-ray burst, taking different time intervals into account along the entire emission. 
In section \ref{sec:3} we present the methods we used to constrain the redshift. 
In section \ref{sec:7}, after interpreting the second episode as a canonical GRB within the fireshell model, we build its light curve and spectrum. 
In Section \ref{sec:8} we go into further detail in the analysis of the first episode, making clear the evolution of the thermal component and the radius of the outermost shell and establishing the complete correspondence with GRB 090618. 
Finally, in Section \ref{sec:9} we present the conclusions.

\section{Brief summary of GRB090618 analysis}\label{GRB090618}

We recall that GRB 0908618 is one of the most energetic among the nearest sources, with an isotropic energy of $E_{iso}= 2.7 \times 10^{53}$ \rm{erg}, at redshift z= 0.54. It has been observed in a wide energy range by many satellites, such as as Fermi GBM \citep{Meegan2}, Swift-BAT \citep{Gehrels2009}, AGILE \citep{Longo2009}, Konus-WIND \citep{Golenteskii2009}, Suzaku-WAM \citep{Kono2009}, and CORONAS-PHOTON \citep{Kotov2008}, and by many onground telescopes. We have shown \citep[see the work of][]{Luca}  that the light curve is quite particular, as it consists of two different emissions, of $50$ s and $100$ s of duration. A time-resolved spectral analysis showed that the first part is well fit by a black body and an extra power-law component. The temperature decays with time following a broken power law, in agreement with the results found by Ryde and collaborators \citep{RydePeer2009}. The first power law has an index $a_{kT}=-0.33 \pm 0.07$, and the second one has an index $b_{kT}=-0.57 \pm 0.11$. The evolution of the radius $r_{em}$ of the black body emitter has also been studied, finding an initial radius of $12000$ km, expanding in the early phase with a velocity of $\sim4000$ km/s. By analyzing it within the fireshell model, we concluded that the first episode cannot be either a GRB or part of a GRB. Indeed, we relate this episode to the phases just preceding the gravitational collapse and define it as a ``proto-black hole'': the latest phase of the collapsing bare core leading to the black hole formation and the simultaneous emission of the GRB \citep{Ruffini2010a}. In this interpretation, the radius $r_{em}$ only depends on the observed energy flux of the black body component $\phi_{obs}$, the temperature $kT$ and the luminosity distance to the source $D$. Episode 2 was identified as a canonical GRB, which comes from the black hole formation process. The first 4 s were identified as the P-GRB, and its spectrum is well fit by a black body with an extra power-law component, the latter mainly caused by the early emission of the extended afterglow. We found a P-GRB temperature of $kT=29.22 \pm 2.21$ \rm{keV} and a dyadosphere energy of the whole second episode of $E^{e^{\pm}}_{tot}=2.49 \times 10^{53}$ \rm{erg}. We performed a numerical simulation with the numerical code GRBsim and found a baryon load $B=(1.98 \pm 0.15) \times 10^{-3}$and a Lorentz Gamma factor at the transparency of $\Gamma= 495 \pm 40$. From this analysis we concluded that we are in the presence of a very interesting source, because for the first time we can witness the process of formation of a black hole from the phases just preceding the gravitational collapse to the GRB emission.

\section{Observations of GRB 101023}\label{sec:2}

On 23 October 2010 the Fermi GBM \citep{GCN11376} detector was triggered by a source quite similar to GRB 090618, with a trigger time of 309567006.726968 (in MET seconds). 
The burst was also detected by BAT \citep{Saxton} (see Fig. \ref{SwiftLC}), onboard the Swift satellite \citep{Gehrels}, with a trigger time of $436981$ (in MET seconds) and the following location coordinates: $RA(J2000) = 21h 11m 49s$, $Dec(J2000) = -65^{\circ} 23' 37''$ with an uncertainty of 3 arcmin. The Swift-XRT detector \citep{GCN11368,Burrows} has also observed this source from 88 s to 6.0 ks after the BAT trigger.  GRB 101023 was also detected by the Wind instrument onboard Konus satellite, in the energy range $(10-770)$ \rm{keV} \citep{GCN 11384}. The inferred location is in complete agreement with that determined by Swift and Fermi. Moreover, there have been detections in the optical band by the Gemini telescope \citep{GCN11366}.

\begin{figure}
\centering
\includegraphics[width=\hsize]{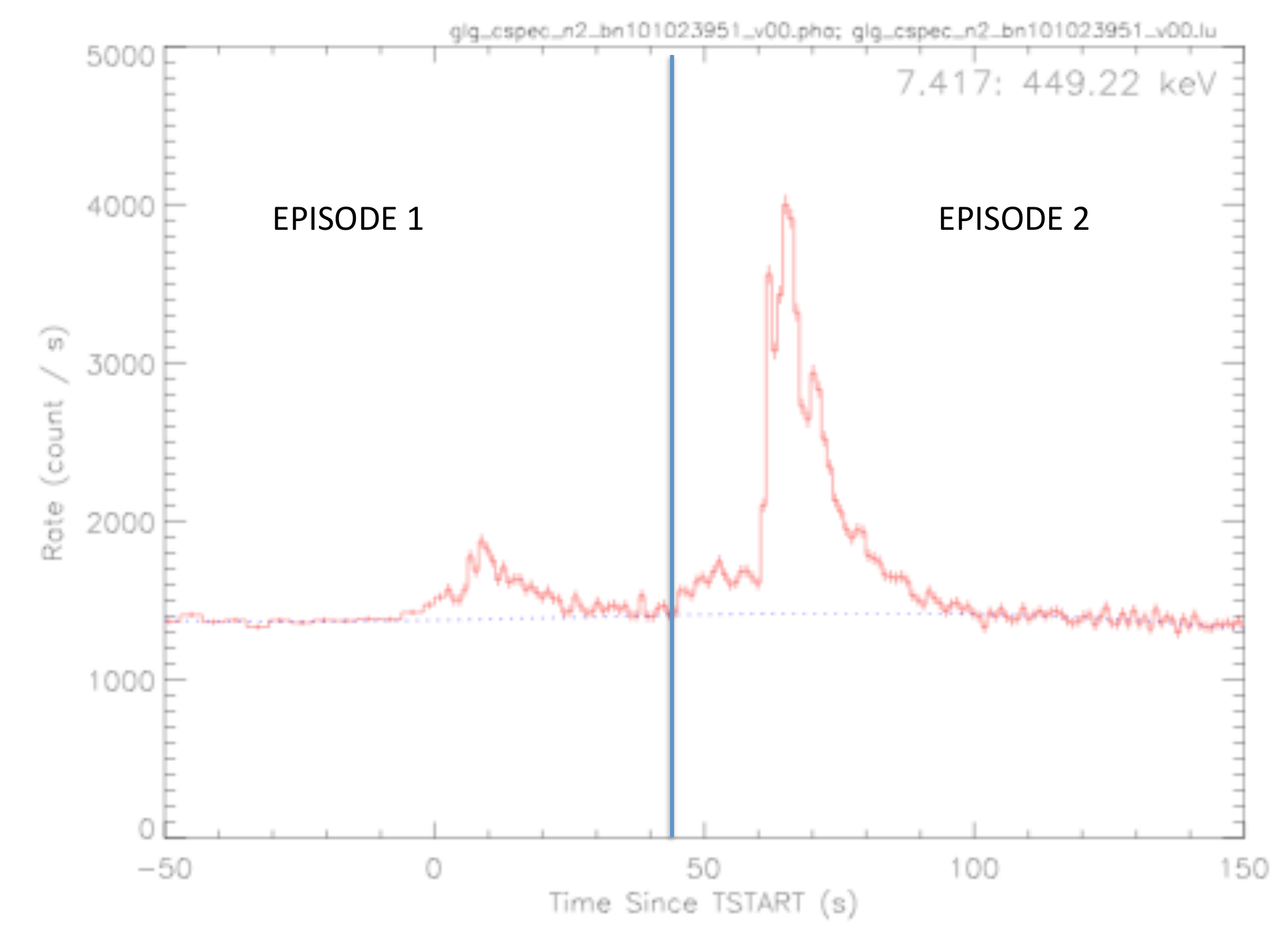}
\caption{Count light curve of GRB 101023 obtained from the Fermi GBM detector, with a bin time of 1 s. The time is given with respect to the GBM trigger time of 22:50:04.73 UT, 2010 October 23. The plot was obtained with the RMFIT program. The two-episode nature of the GRB is shown in analogy with GRB 090618.}
\label{picture of the light curve}
\end{figure}

\begin{figure}
\centering
\includegraphics[width=\hsize]{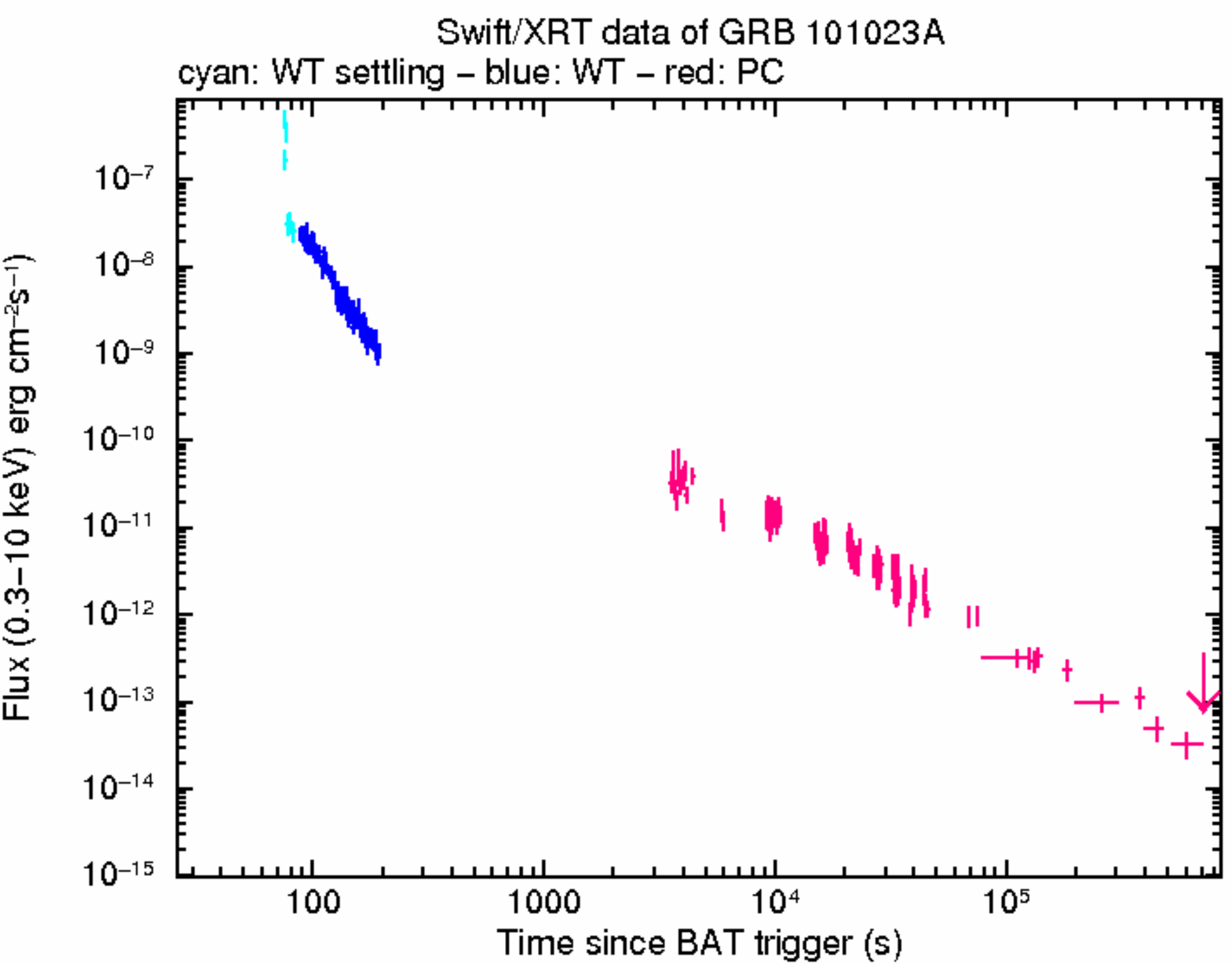}
\caption{Count light curve of GRB 101023 obtained from the Swift XRT detector.}
\label{SwiftLC}
\end{figure}

The GBM light curve (Fig. \ref{picture of the light curve}) shows two major pulses.The first one starts at the trigger time and lasts 45 s. It consists of a small peak that lasts about 10 s, followed by a higher emission that decays slowly with time. The duration, as well as the topology of this curve, lead us to think that this may not be a canonical GRB, but its origin may lie on another kind of source, which remains unidentified.  The second pulse starts at 45 s after the trigger time and lasts 44 s. It presents a peaky structure, composed of a short and weak peak at the beginning, followed by several bumps, big not only in magnitude but also in duration. This second emission, in contrast, does have all the characteristics that describe a canonical GRB \citep{GCN11459}.

\section{Theoretical model considered: Fireshell Scenario}\label{sec:4}

In the fireshell scenario, the GRB emission comes from a process of vacuum polarization, resulting in pair creation in the so-called dyadosphere. In the process of gravitational collapse to a black hole \citep{Ruffini2010b}, an $e^{\pm}$ plasma is formed in thermal equilibrium, with total energy $E^{e^{\pm}}_{tot}$.  The annihilation of these $e^{\pm}$ pairs occurs gradually and is confined in a shell, called ``fireshell''. This shell self-accelerates to relativistic velocities, engulfing the baryonic matter (of mass $M_B$) left over in the process of collapse and reaching a thermal equilibrium with it \citep{Ruffini2000}. The baryon loading is measured by the dimensionless parameter $B=M_B c^2 / E^{e^{\pm}}_{tot}$. The fireshell continues to self-accelerate up to relativistic velocities \citep{Ruffini1999} until it reaches the transparency condition. At this time we have a first flash of radiation, the P-GRB \citep{Ruffini2001}. The energy released in the P-GRB is a fraction of the initial energy of the dyadosphere \rm{$E^{e^{\pm}}_{tot}$}. The residual plasma of leptons and baryons interacts with the circumburst medium (CBM) as it expands, giving rise to multi-wavelength emission: the ``extended'' afterglow. However, owing to these collisions, the plasma starts to slow down. We assume a fully-radiative condition in this model \citep{Ruffini2003}. The structures observed in the prompt emission of a GRB come from the inhomogeneities in this CBM, while in the standard fireball scenario \citep{Meszaros2006} they are caused by internal shocks. In this way we need few parameters for a complete description of a GRB: the dyadosphere energy \rm{$E^{e^{\pm}}_{tot}$}, the baryon load $B$ and the CBM density distribution, $n_{CBM}$.  In addition, we assume that there is spherical symmetry, and the energy released in the explosion $E_{iso}$ is equal to the energy of the dyadosphere $E^{e^{\pm}}_{tot}$. From this approach, to sum up, the GRB bolometric light curve will be composed of two main parts: the P-GRB and the extended afterglow. Their relative energetics and their observed time separation are functions of the parameters  $E^{e{\pm}}_{tot}$, $B$, and $n_{CBM}$. We want to stress that the emission of the P-GRB does not always coincide with what is called ``prompt emission'' in the fireball scenario. Indeed, within the fireshell model, this prompt emission corresponds to the gamma-ray emission, which addresses not only the P-GRB, but also the peak of the extended afterglow.

Instead of making use of the typical thermal spectrum, we introduced a modified black body spectrum \citep{Patricelli2010,Patricelli2011}, given by

\begin{equation}
\frac{dN_{\gamma}}{dVd{\epsilon}} = \left(\frac{8\pi}{h^3c^3}\right) \left(\frac{\epsilon}{k_BT}\right)^{\alpha} \frac{\epsilon^2}{exp \left(\frac{\epsilon}{k_BT}\right) -1}.
\label{spettro_barbara}
\end{equation}
This way we can also reach an agreement with the most energetic GRBs ($E_{iso} \geq 10^{53}$ \rm{erg}). 
Furthermore, within the fireshell scenario we can naturally explain the hard-to-soft spectral variation observed in the extended afterglow emission. As the Lorentz gamma factor $\Gamma$ decreases with time, the observed effective temperature of the fireshell also decreases, making the peak of the emission take place at lower energies. This effect is amplified by the curvature effect of the EQTS \citep{Bianco}, which produces the observed time lag in the majority of the GRBs.

We need to identify the P-GRB in the observed data so that we are able to determine the parameters $E^{e^{\pm}}_{tot}$ and $B$, via a trial and error procedure, and consequently the P-GRB energy $E_{P-GRB}$, the Lorentz gamma factor at the transparency $\gamma$, the theoretically predicted temperature $kT_{th}$, and the radius at the transparency (see Fig. 1 in \citep{Ruffini2009}).
The observed temperature $kT_{obs}$ is related to the theoretically predicted temperature $kT_{th}$ through

\begin{equation}
\label{Doppler}
 kT_{obs}=\frac{kT_{th}}{1+z}.
\end{equation}

\section{Analysis of data and results}\label{sec:5}

To obtain the Fermi GBM light curve and spectrum in the band $8-440$  \rm{keV} (see Fig. \ref{picture of the light curve}), we used the RMFIT program. We downloaded the data from the gsfc website\footnote{ftp://legacy.gsfc.nasa.gov/fermi/data/gbm/bursts}. We used the lightcurves corresponding to the second and fifth NaI detectors and the b0 BGO detector. We subtracted the background by fitting a cubic function from the intervals before and after the GRB (from 400 s to 200 s before the GRB and from 180 s to 220 s after it), where we suppose there is no data. Then we proceeded with the time-resolved spectral analysis. 

\begin{figure}
\centering
\includegraphics[width=\hsize]{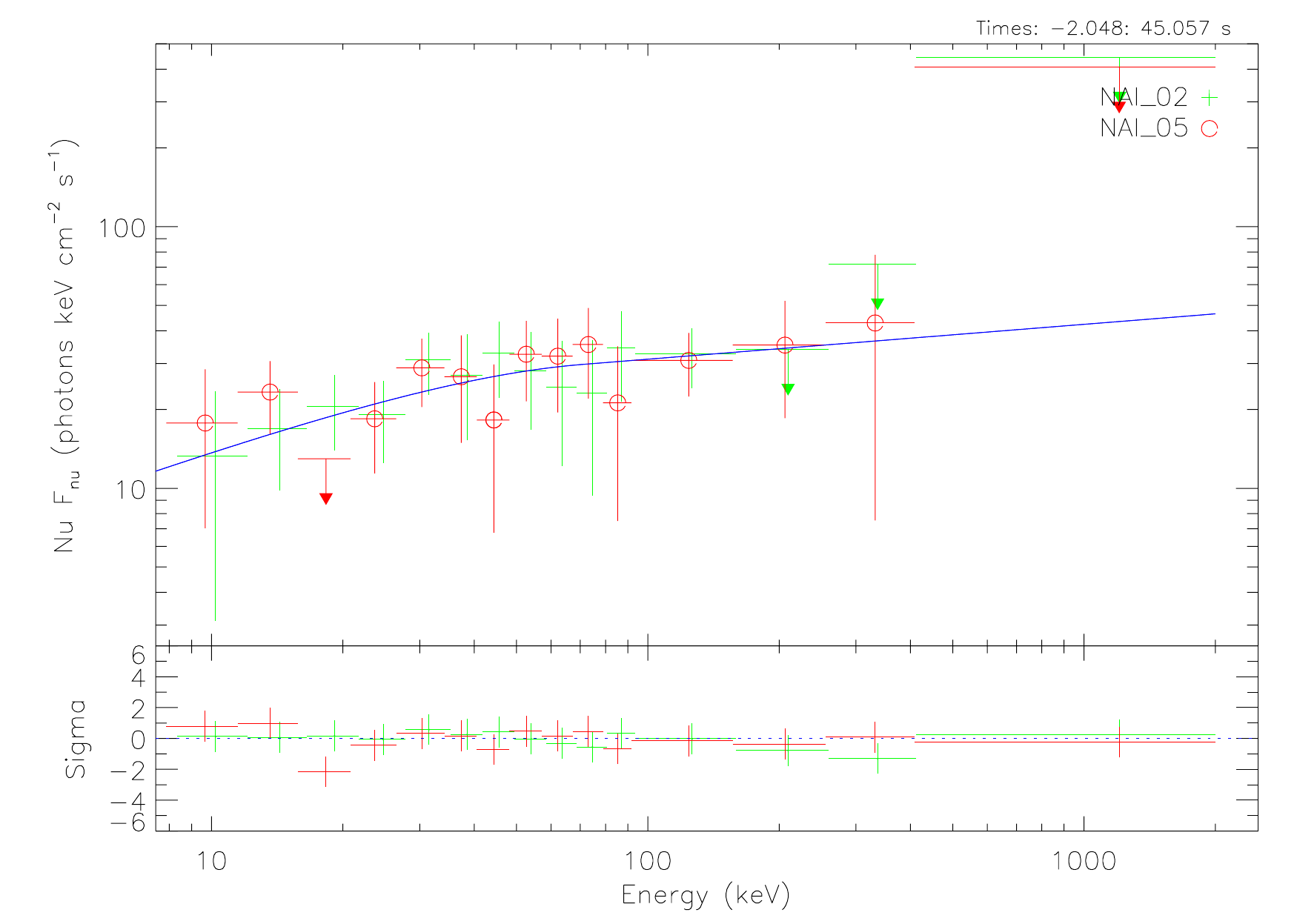}
\includegraphics[width=\hsize, height= 7cm]{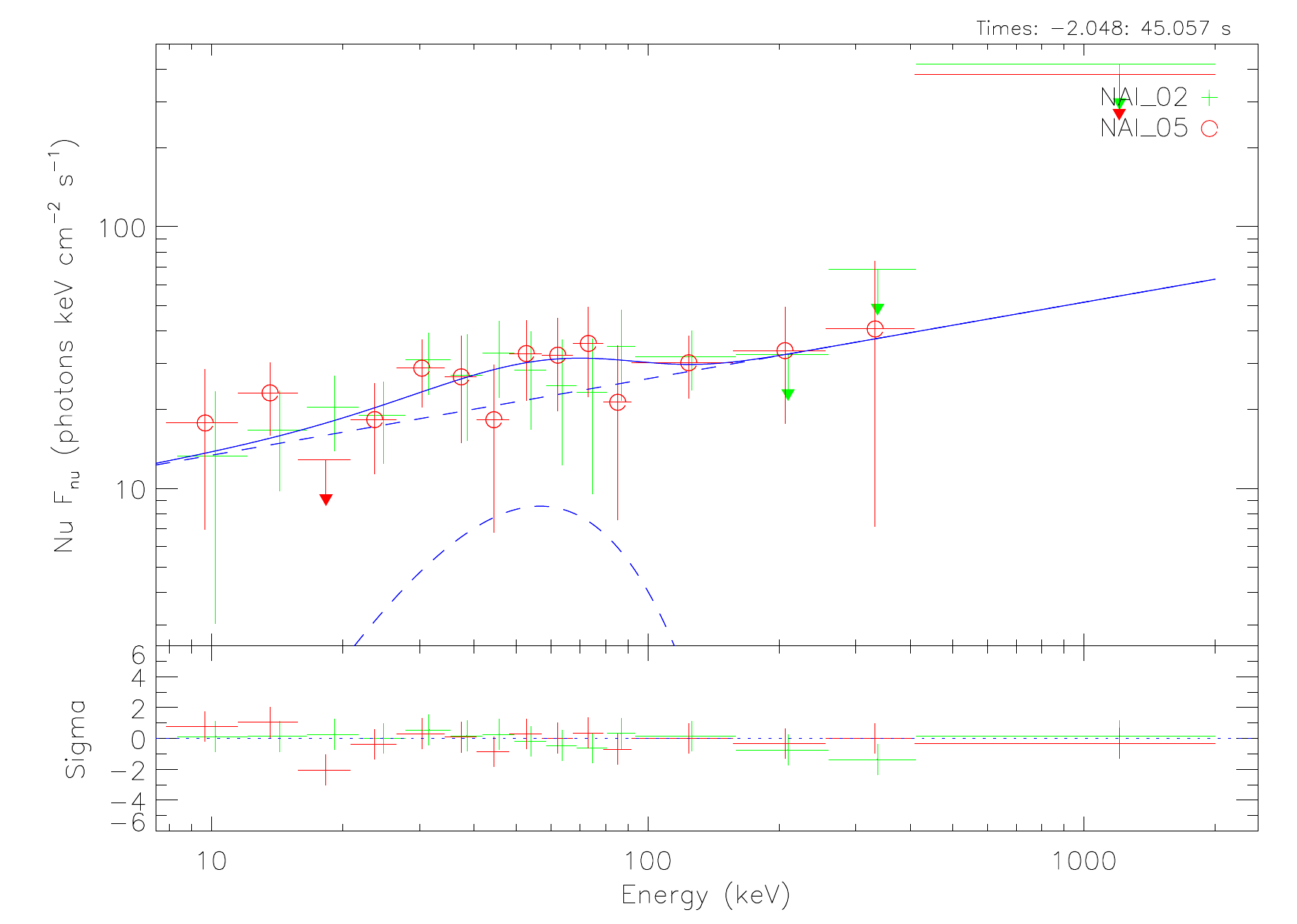}
\caption{Fit of the spectrum of episode 1, with a Band model (upper panel) and a black body plus power-law  model (lower panel).}
\label{spettroGRB1}
\end{figure}

\begin{figure}
\centering
\includegraphics[width=\hsize]{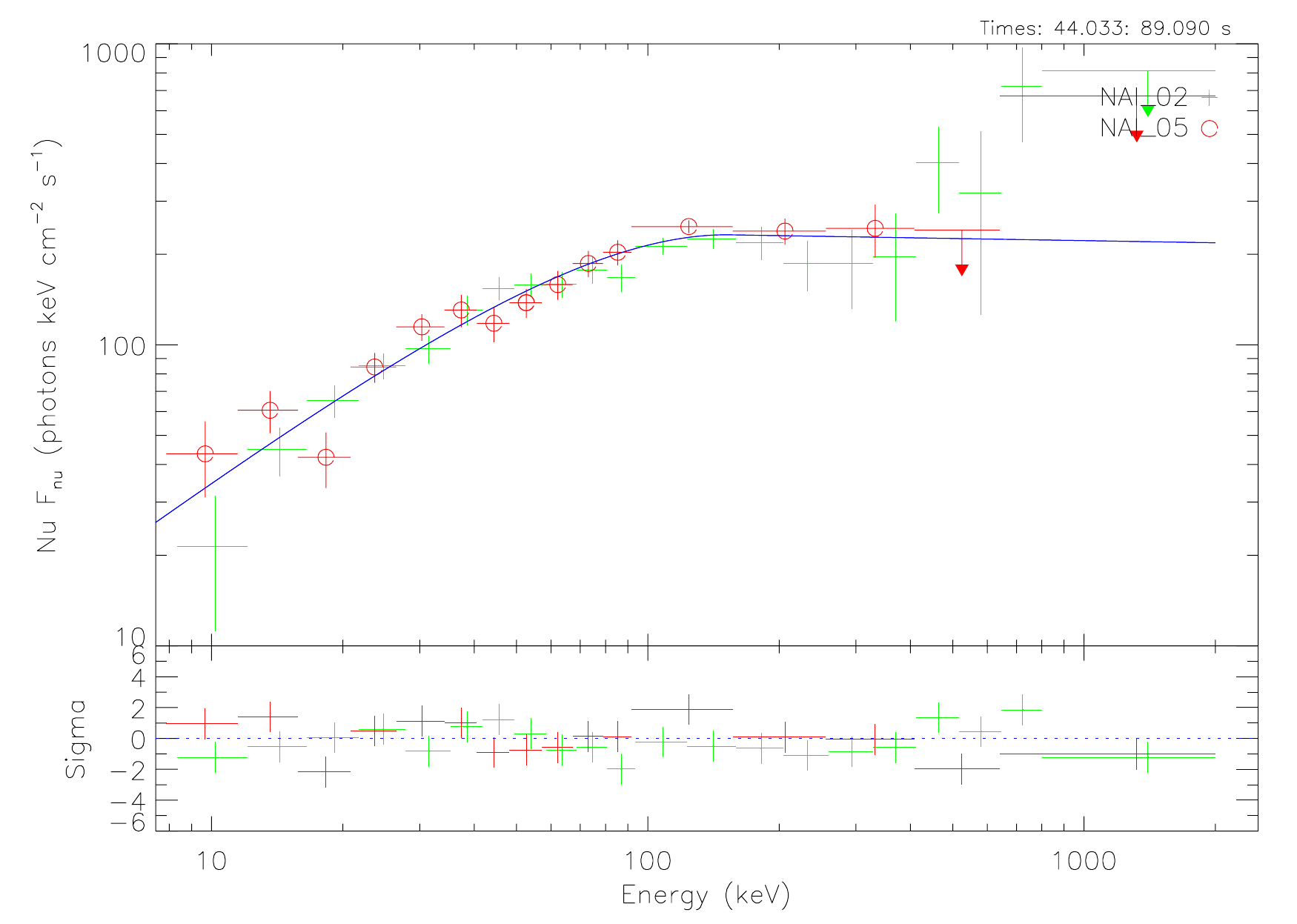}
\includegraphics[width=\hsize]{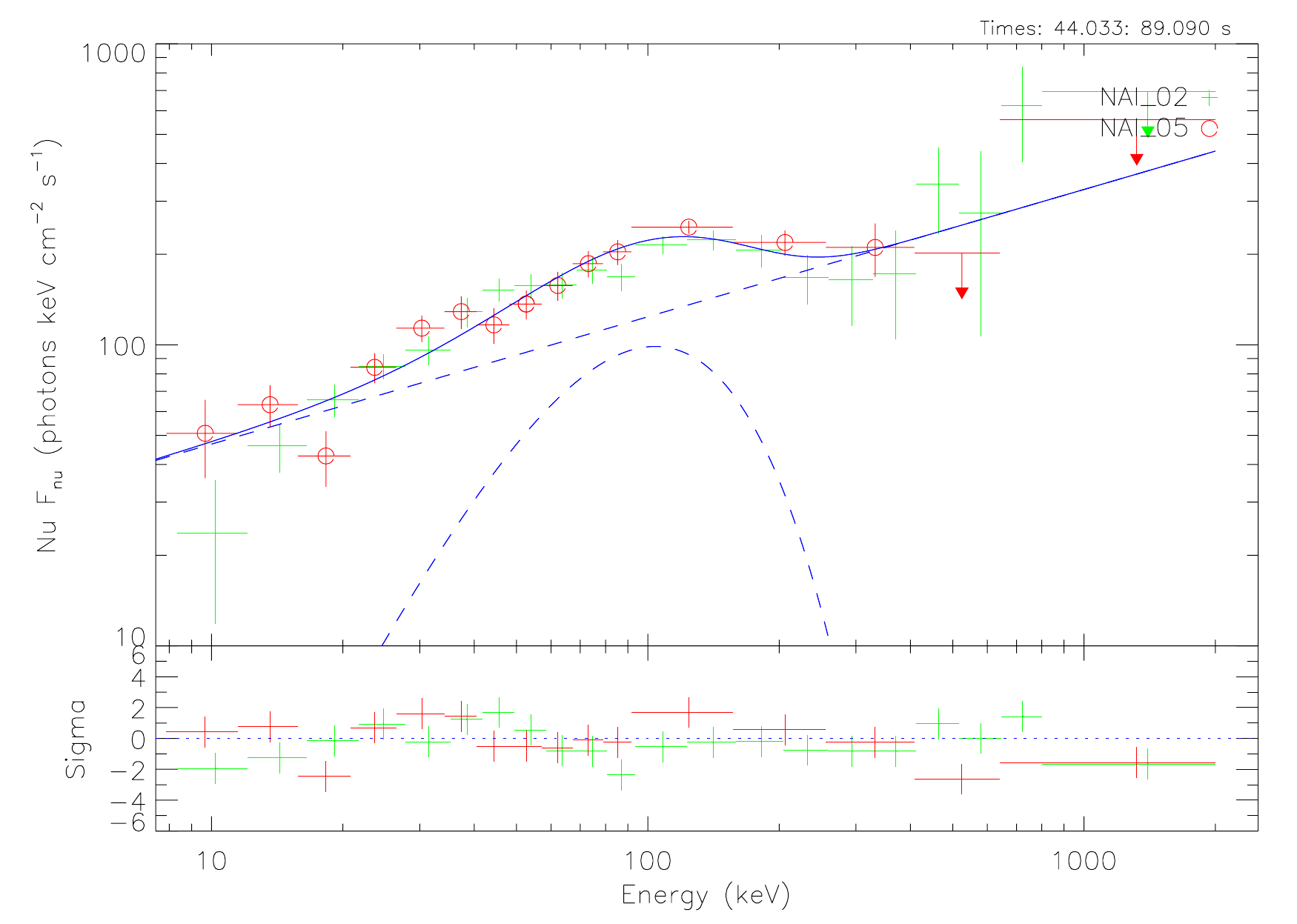}
\caption{Fit of the spectrum of episode 2, with a Band model (upper panel) and a black body plus a power-law model (lower panel). Both models fit the entire energy range well, with a chi squared of 0.79 and 0.84, respectively. The data points have been grouped according a signal-to-noise ratio of N=10, and rebinned at higher energies in order to have better statistics and reduce the error bars.}
\label{spettroGRB2}
\end{figure}

To proceed with the fitting of the spectra, we defined first of all the time intervals we wanted to analyze: the first interval starts at the trigger time $t_0=0$ and lasts 45 s, while the other starts at $t_0 + 45$ s and lasts 44 s. For convenience, from now on we will refer to the first emission as episode 1 and the second emission as episode 2. For this source we considered two models: the black body plus power-law model and the Band spectral model \citep{Band}. We first analyzed each of the events separately, as if they were two GRBs and then subdivided each of the two emissions in the light curve into two other parts: the one that we think would correspond to the P-GRB emission and the one that would correspond to the afterglow. The results from the spectral analysis are shown in Table \ref{TABLA 1}. The fit of the spectrum of the first episode with both models is shown in Fig. \ref{spettroGRB1}, while Fig. \ref{spettroGRB2} shows the same fit for the second episode.

\begin{table*}
\centering
\caption{Time-resolved spectral analysis of GRB 101023.} 
\label{TABLA 1} 
\begin{tabular}{l c c c c c| c c c c c }
\hline\hline
Time $[s]$& $\alpha$        & $\beta$                & $E_0^{BAND}[\rm{keV}]$       & $\chi^2$ & Norm                   &$kT [\rm{keV}]$         &$\gamma$      &$\chi^2$& Norm $^{po}$ & Norm $^{BB}$\\
0-44                      & -1.3$\pm$0.8& -1.9$\pm$0.2       & 87$\pm$147           & 0.98        & 0.006$\pm$0.01&14$\pm$6 & -1.7$\pm$0.1    &0.98      &0.0003$\pm$0.0004 & (4.1 $\pm7.4) \times 10^{-5}$\\
45-89                    &-0.9$\pm$0.1 & -2.02$\pm$0.1     &151$\pm$24            & 1.09        &0.043$\pm$0.008&26$\pm$1&-1.58$\pm$0.03&1.12   &0.0124$\pm$0.0006& (4.2 $\pm1.1)\times 10^{-5}$\\
\hline
\hline
\end{tabular}
\end{table*}

\section{Identification of the P-GRB}\label{sec:6}

\subsection{Attempt for a single GRB scenario: the whole emission as a single GRB}

The first step in our analysis was to attempt to interpret the whole emission as a single GRB, with episode 1 as the P-GRB. We performed a time-integrated analysis of the whole emission of episode 1, using a black body plus power-law model and a Band model. The results of this spectral analysis are shown in Table \ref{TABLA 1}. We found a black-body temperature of $kT=14 \pm 6$ \rm{keV} with normalization factor $norm_{bbody}=(4.1 \pm 7.4) \times 10^{-5}  $, a photon index of $\gamma=-1.7 \pm 0.1$ with normalization factor $norm_{po}=(3 \pm 4) \times 10^{-4}  $ and a $\chi^2=0.98$ for both spectral models. The P-GRB energy is $E_{P-GRB}=1.625 \times 10^{52}$ erg and the isotropic energy $E_{iso}=4.03 \times 10^{53}$ erg, which gives a ratio $E_{P-GRB}/E_{iso}= 0.04$. This value in our simulations would imply a theoretically predicted temperature of $kT_{th}=110.63$ keV, which is by far much bigger than the observed one. Consequently, the first episode cannot be the P-GRB of the whole emission.

\subsection{The identification of the P-GRB of the first episode}

Our second step in the analysis of this source was to attempt to interpret episodes 1 and 2 as two different GRBs. We first analyzed episode 1 by taking two different possibilities into account: 

1) We considered a P-GRB that lasts 6 s and made the spectral analysis with XSPEC. We fitted a black body plus power-law model and found a black-body temperature of $kT=25.4 \pm 6.9$ \rm{keV} with normalization factor $norm_{bbody}=0.9 \pm 0.5  $, a photon index of $\gamma=2.2 \pm 0.5$ with $norm_{po}=30.9 \pm 35.3  $ and a reduced chi squared of $\chi^2=1.01$. Considering that the P-GRB is the thermal component of the GRB, by using XSPEC we found a flux of $7.25 \times 10^{-8} \rm{erg}/cm^{2}/s$ in the range (8-5000) \rm{keV}. Then we followed the same procedure for the whole of episode 1, fitting a cutoffpl model, and found a photon index of $\gamma= 1.16 \pm 0.3$, a cutoff energy of $E_{cutoff}=73 \pm 27$ \rm{keV}, a normalization factor of $2.9 \pm 2.4 $, a reduced chi squared value of $\chi^2=1.08$, and a flux of $1.626 \times 10 ^{-7} \rm{erg}/cm^{2}/s$. Using formula \ref{Eiso}, we found a P-GRB energy of $E_{P-GRB}=9.56 \times 10^{50}$ \rm{erg} and a total energy of $E^{e^{\pm}}_{tot}=1.625 \times 10^{52}$ \rm{erg}, which gives a ratio $E_{P-GRB}=5.9\%E^{e^{\pm}}_{tot}$. With these values we performed the simulation with the numerical code and found a baryon load $B=8.5 \times 10^{-4}$ and a predicted temperature of $kT_{th}=128.82$ \rm{keV}, which is much higher than the one observed. Therefore, we concluded that the first 6 s of emission cannot be the P-GRB of episode 1, at least in the fireshell scenario. 2) We considered the P-GRB under the threshold of the detector. We took the first 6 s before the trigger time as the P-GRB and supposed that it is well fitted by a Band model, with a flux of $10^{-8} \rm{erg}/ cm^{2}/s$, which is comparable with the threshold of the detector. We derived a P-GRB energy of $10^{50}$ \rm{erg}, which is the $0.9\%$ of the total energy. For this ratio of the energies, we found with the numerical code a baryon load of $B=10^{-2}$ and a predicted flux that is smaller than the detector threshold. This indicates that indeed this could be the P-GRB of the first emission, so that episode 1 could be a GRB, and we could be for the first time in the presence of a double GRB. However, in light of the results obtained from the analysis of GRB 090618 \citep{Luca} and taking into account that the value of the redshift has not been precisely determined, we decided to discard this result. Therefore, we conclude that episode 1 is not a GRB but another source whose origin is still unidentified. We come back to this interpretation later.

\subsection{Analysis of the second episode}

After the analysis of episode 1, we moved on to the analysis of episode 2. We followed the same steps taking the first 12 s of episode 2 as the possible P-GRB. We also fitted a black body plus power-law model to the whole P-GRB and found a black-body temperature of $kT=15.5 \pm 1.6$ \rm{keV} with normalization factor $norm_{bbody}=1.26 \pm 0.3  $, a photon index of $\gamma=2.5 \pm 0.4$ with normalization factor $norm_{po}=141.79  $ and a $\chi^2=0.96$. We computed a flux in the band (260-5000) \rm{keV} of $2.54 \times 10^{-7} \rm{erg}/cm^2/s$ and a P-GRB energy of $E_{P-GRB}=1.89 \times 10^{52}$ \rm{erg}. By fitting a black body plus power-law model to the whole of episode 2 we found a flux in the band (8-5000) \rm{keV} of $1.272 \times 10^{-7} \rm{erg}/cm^2/s$ and a total energy of $E^{e^{\pm}}_{tot}=1.309 \times 10^{53}$ \rm{erg}. The ratio is $E_{P-GRB}=0.9 \%E^{e^{\pm}}_{tot}$. This same value is reached with the numerical code for a baryon load $B=7.6 \times 10^{-3}$ and a predicted temperature of $kT_{th}=14.02$ \rm{keV}, which after cosmological correction gives 7.38 \rm{keV} (assuming z=0.9, see next section), which is not in good agreement with the observed one, $kT_{obs}= 26$ \rm{keV}. Thus we conclude that the first 12 s of emission cannot be the P-GRB. 

\begin{figure}
\centering
\includegraphics[width=\hsize]{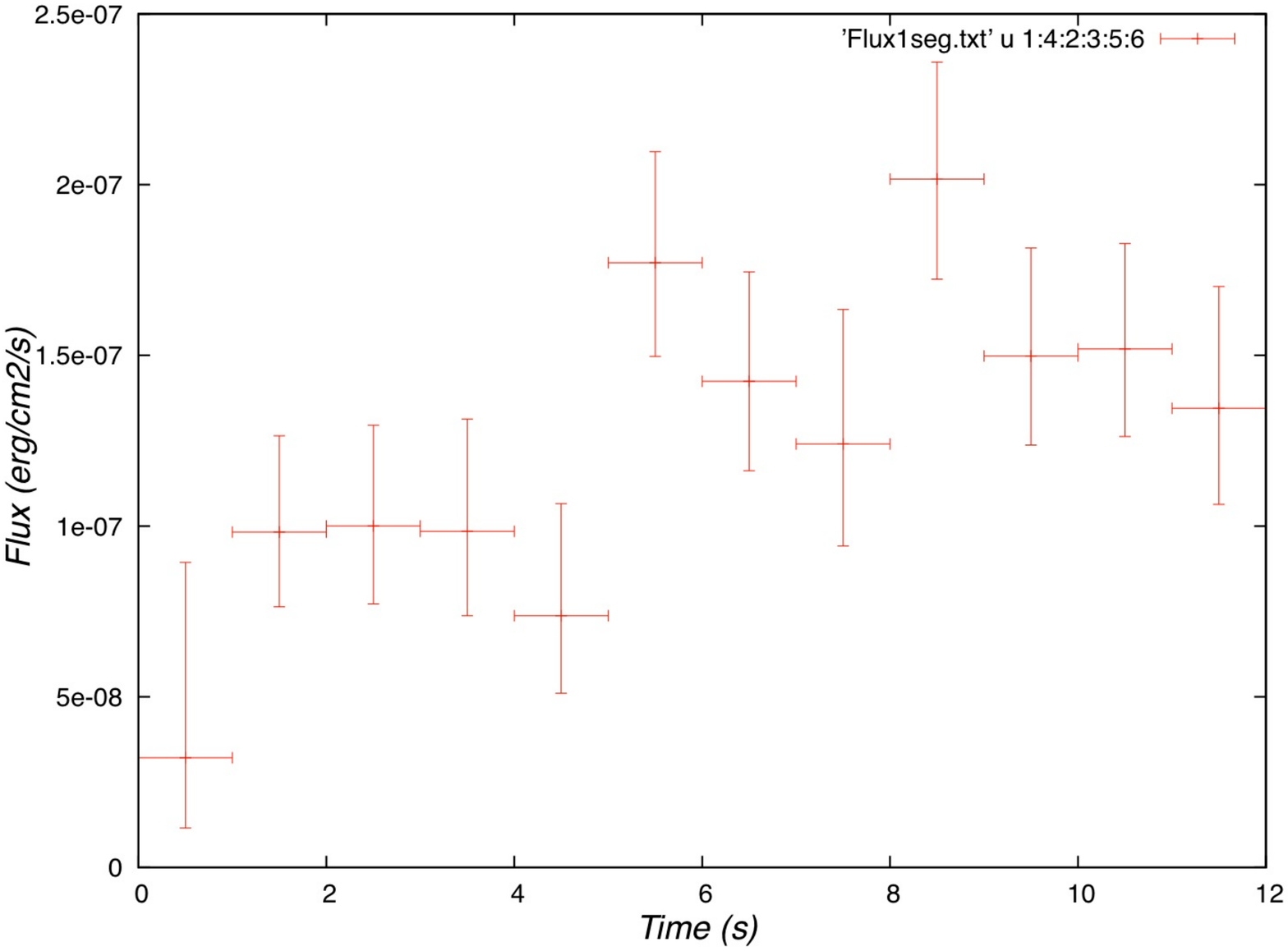}
\caption{Plot of the flux of the BB component vs time for the first 12 s of episode 2.}
\label{Flux_1seg_GRB2}
\end{figure}

To be more accurate, we performed the following procedure: as we know that the P-GRB consists of a black-body emission, we performed a detailed spectral analysis every 1s with the Black body model to see the behavior of the black body component, i.e where the black body component dominates. That will indicate more precisely the time range and duration of the P-GRB. Table \ref{TABLA 3} shows the results of this analysis and Fig. \ref{Flux_1seg_GRB2} shows the behavior of the black body component with time. We see that in fact only the first 5 s of emission have a marked black body component, with a typical pulse shape. The emission that follows seems not to be related to the P-GRB, but to the afterglow. So we conclude that episode 2 is indeed a GRB and the first 5 s of emission are the P-GRB (see Section \ref{sec:7}).

\begin{table*}
\centering
\caption{Detailed spectral analysis of the P-GRB of episode 2, of 12s of duration, using a BB+po model and performed every 1 second.}
\label{TABLA 3}
\begin{tabular}{ccccc}
\hline
P-GRB of episode 2& (BB+po)&&&\\
\hline
Time Int & kT(\rm{keV})       & norm factor  &  Flux 8-440 keV             &$\chi^2$\\
\hline
051-052&1.9$\pm$1.7 & 0.9$\pm$2   &  3.2525 $\times 10^{-8}$& 1.40\\ 
052-053&5$\pm$1       & 1.3$\pm$0.3&  9.8254$\times 10^{-8}$ & 1.06\\ 
053-054&7$\pm$1       & 1.2$\pm$0.3&  9.9689$\times 10^{-8}$&  0.99\\
054-055& 10$\pm$2    & 1.2$\pm$0.3 &  9.8285$\times 10^{-8}$ & 1.17\\
055-056& 7$\pm$1      & 1.6$\pm$0.3 & 1.3217$\times 10^{-7}$ & 0.96\\  
056-057& 10$\pm$1    & 2.1$\pm$0.4  &1.7721$\times 10^{-7}$ & 1.42\\
057-058&10$\pm$1     &1.7$\pm$0.4   & 1.4245$\times 10^{-7}$ & 0.96\\
058-059& 11$\pm$1    & 2.1$\pm$0.4  & 1.7738$\times 10^{-7}$ & 1.16\\
059-060& 10$\pm$1    & 2.6$\pm$0.4  & 2.1844$\times 10^{-7}$ & 1.38\\
060-061& 10$\pm$1    & 1.8$\pm$0.3  & 1.4976$\times 10^{-7}$  & 1.51\\
061-062& 9$\pm$1      & 1.8$\pm$0.3  & 1.5193$\times 10^{-7}$ & 1.18\\
062-063&14$\pm$2      &1.6$\pm$0.4  & 1.3462$\times 10^{-7}$ & 1.74\\
\end{tabular}
\end{table*}

\section{Pseudo-redshift determination}\label{sec:3}

The redshift of this source is unknown, owing to the lack of data in the optical band. However, to constrain it, we employed three different methods, mentioned below.

\subsection{Method 1: nH column density}

We first tried to estimate the redshift making use of the  method developed in \citet{Grupe} work, where the authors comment on the possible relation between the absorption column density in excess of the galactic absorption column density $\Delta N_H = N_{H,fit} - N_{H,gal}$ and the redshift z. 
To do this, we considered the galactic absorption component taken from \citet{Kalberla2005}, with the following values of the galactic coordinates of the GRB:  $l = 328.88$, $b = -38.88$. We used the Lab Survey website\footnote{http://www.astro.uni-bonn.de/$\sim$webaiub/english/tools$\_$labsurvey.php} and obtained the value of $n_H=2.59 \times 10 ^{20} cm^{-2}$ for the galactic H column density. 

Then we took the values of some parameters, the spectrum, and response files from the XRT website\footnote{http://www.swift.ac.uk/xrt$\_$curves/}, selected the part of interest and carried out an analysis making use of the program XSPEC. We fit the model wabs, which is the photoelectric absorption using Wisconsin cross sections \citep{Wabs}: $M(E)= exp[-n_H \sigma(E)]$, where $\sigma(E)$ is the photoelectric cross section (not including Thomson scattering) and $n_H$ is the equivalent hydrogen column density, in units of $10^{22}$ atoms/cm$^{2}$. 
Once we knew these parameters, we fit the data with a power-law model, considering a phabs component related to the intrinsic absorption.
We obtained a value of $n_H^{intr}= 0.18 \pm 0.019 \times 10^{22} cm^{-2}$. Wkth this result, we put them in formula (1) of \citet{Grupe} paper:
\begin{equation}
log(1+z) < 1.3-0.5[log(1+\Delta N_H)],
\end{equation}
and we obtained an upper limit for the redshift of 3.8.\\

\subsection{Method 2: Amati relation}
 
\begin{figure}
\centering
\includegraphics[width=\hsize]{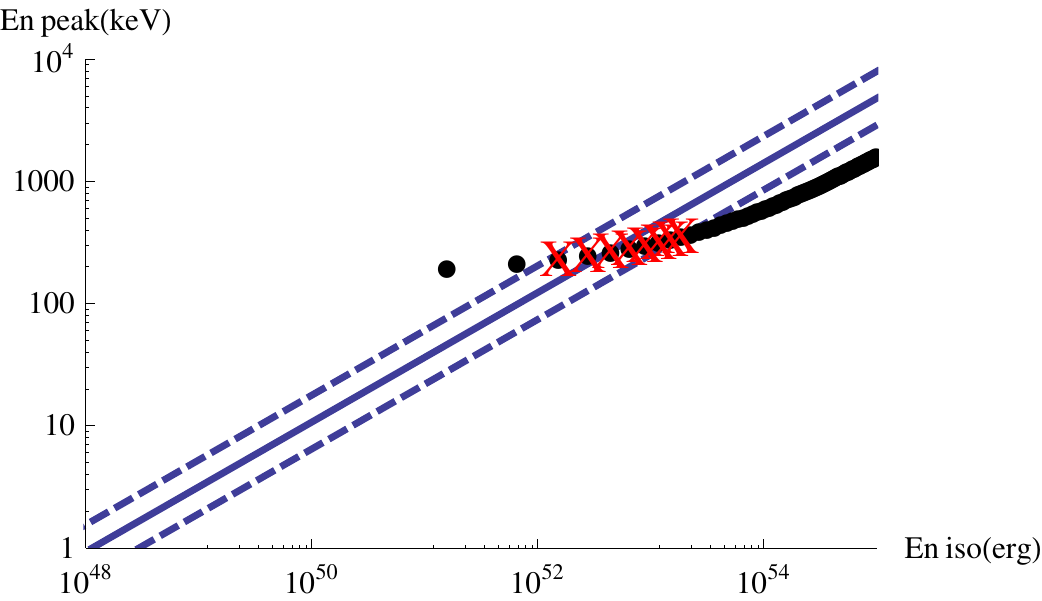}
\caption{Plot of the relation between $E_{p,i}$ and $E_{iso}$ for the second episode of GRB 101023, considering different values of the redshift. It can be seen that the plot lies within $1\sigma$ for the range z= 0.3 - z= 1.0.}
\label{PICTURE OF AMATI RELATION}
\end{figure}

We tried another method of constraining the redshift, making use of the Amati relation \citep{Amati(2006)}, shown in Fig. \ref{PICTURE OF AMATI RELATION}.  This relates the isotropic energy $E_{iso}$ emitted by a GRB to the peak energy in the rest frame $E_{p,i}$ of its $\nu F_{\nu}$ electromagnetic spectrum \citep[see][and references therein]{Amati(2009)}. $E_{iso}$ is the isotropic-equivalent radiated energy, while $E_{p,i}$ is the photon energy at which the time averaged $\nu F_{\nu}$ spectrum peaks. The analytical expression of $E_{iso}$ is
\begin{equation}
\label{Eiso}
E_{iso}=\frac{4 \pi d_l^2}{(1+z)} S_{bol},
\end{equation}
where $d_l^2$ is the luminosity distance, $z$ is the redshift and $S_{bol}$ is the bolometric fluence, related to the observed fluence in a given detection band ($E_{min}$, $E_{max}$) by
\begin{equation}
S_{bol}=S_{obs}\frac{\int^{10^4 / (1+z)}_{1/ (1+z)} E \phi(E) dE}{\int ^{E^{max}}_{E^{min}} E \phi(E) dE},
\end{equation}
with $\phi$ the spectral model considered for the spectral data fit. The value of $E_{p,i}$ is related to the peak energy $E_p$ in the observer's frame by
\begin{equation}
E_{p,i}=E_p (1+z)\, .
\end{equation}

We started our analysis under the hypothesis that episode 2 is a long GRB. We computed the values of $E_{p,i}$ and $E_{iso}$ for different given values of $z$ and plotted them in Fig. \ref{PICTURE OF AMATI RELATION}. We found that the Amati relation is fulfilled by episode 2 for $0.3 < z < 1.0$. This interval has been calculated at $1 \sigma$ from the best fit from the Amati relation, in order to obtain a tighter interval around the best fit than with the previous method.

\subsection{Method 3: Empirical method for the pseudo-redshift}

We also tried an empirical method, following \citet{Atteia} and \citet{Pelangeon}, which can be used as a redshift indicator. This method consists in determining a pseudo-redshift from the GRB spectral properties. Using the parameters from the Band model, namely the index of the low-energy power-law $\alpha$ and the break energy $E_0$, we can compute the value of the peak energy of the $\nu F_{\nu}$ spectrum, as $E_p=E_0(2+\alpha)$. Then, we define the isotropic-equivalent number of photons in a GRB, $N_{\gamma}$, as the number of photons below the break, integrated from $E_p/100$ to $E_p/2$. If we also know the $T_{90}$, we define the redshift indicator
\begin{equation}\label{X}
X=\frac{N_{\gamma}}{E_p \sqrt{T_{90}}}.
\end{equation}

From a sample of 17 GRBs with known redshift reported in \citet{Atteia} we compute the theoretical evolution of X with the redshift z, that is $X=f(z)$. Then we invert the relation to derive a pseudo-redshift from the value of X. That way we obtain the pseudo-redshift as $\hat{z}=f^{-1}(X)$, for the GRB of interest. 

We applied this treatment to episode 2 of GRB 101023, introducing the spectral parameters from the Band model on the Cosmos website\footnote{http://cosmos.ast.obs-mip.fr/projet/v2/fast\_computation.html} and obtained a value for the redshift of $z = 0.9 \pm 0.084$. It is important to mention here that this error is a statistical one, while the systematic error is much bigger \citep{Atteia, Pelangeon,Pelangeon2008}, of a factor of $\sim$ 1.5, i.e., $z= 0.9_{-0.3}^{+0.45}$. 

This result agrees with the redshift range found from the Amati relation for episode 2 and is also consistent with the upper limit determined with method 1.

\section{Simulation of the light curve and spectrum}\label{sec:7}

\begin{figure}
\centering
\includegraphics[width=\hsize]{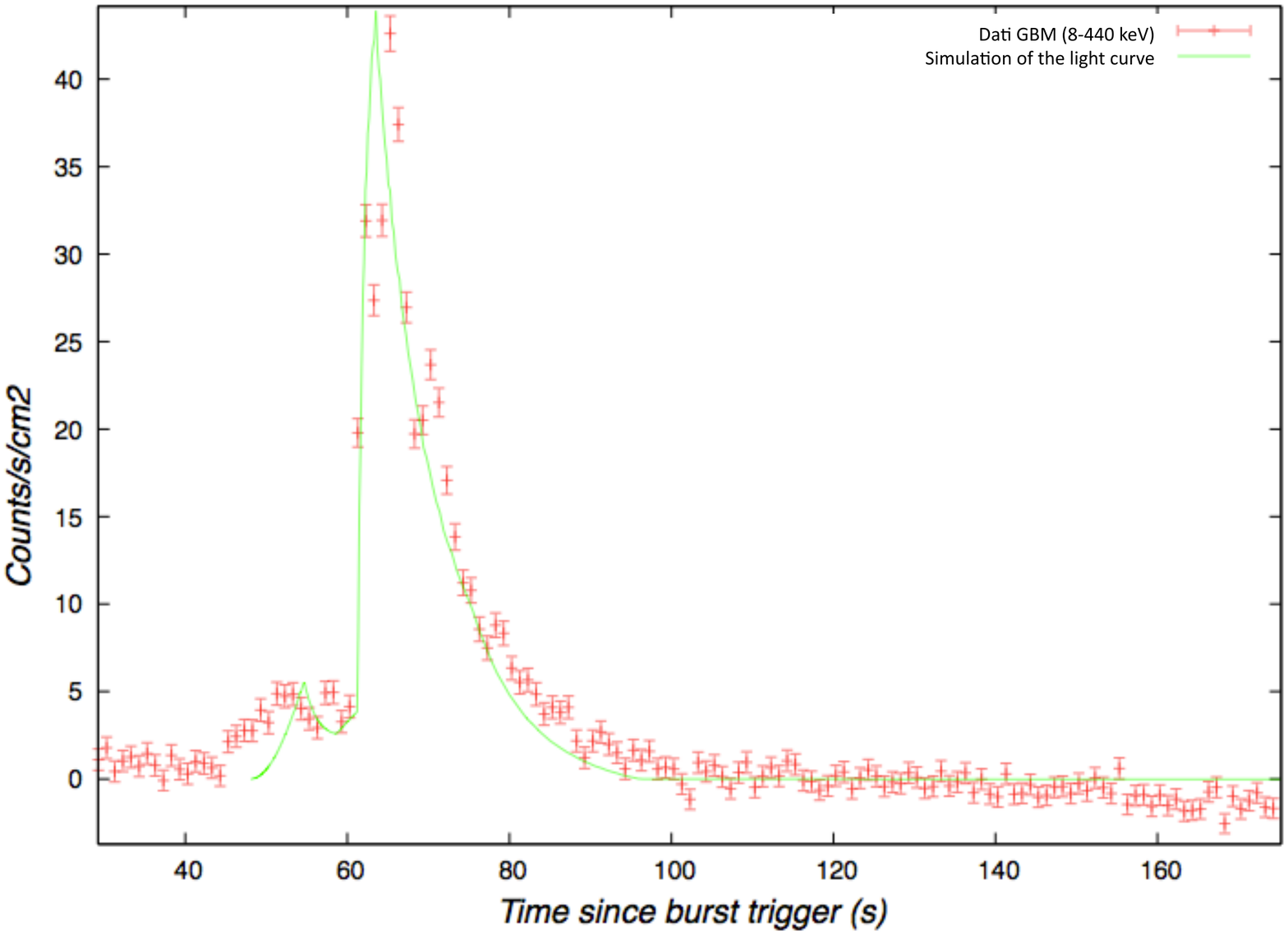}
\caption{Fit of the second major pulse of the light curve of GRB 101023.}
\label{FITTED LIGHT CURVE 2}
\end{figure}

\begin{figure}
\centering
\includegraphics[angle=-90, width=\hsize]{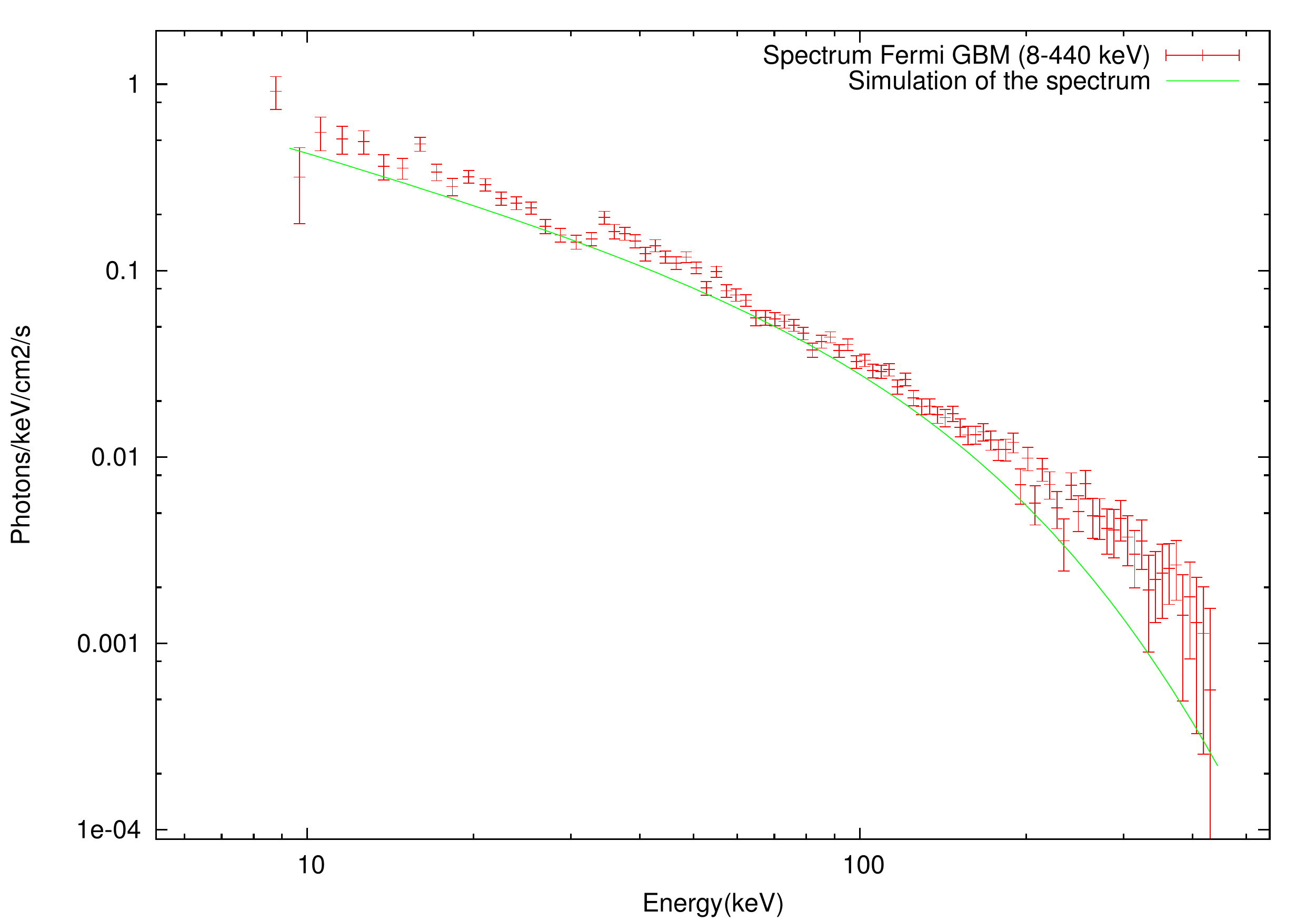}
\caption{Fit of the spectrum of episode 2.}
\label{FITTED SPECTRUM}
\end{figure}

To simulate the light curve we made use of a numerical code called GRBsim. This numerical code simulates a GRB emission by solving the fireshell equations of motion, taking the effect of the EQuiTemporal Surfaces \citep[EQTS,][]{Bianco} into account. We made the simulation for episode 2. We found, at the transparency point, a value of the laboratory radius of $1.34 \times 10^{14} cm$, a theoretically predicted temperature that after cosmological correction gives $kT_{th}=13.26$ keV, a Lorentz Gamma factor of $\Gamma= 260.48$, a P-GRB laboratory energy of $2.51 \times 10^{51}$ \rm{erg} and a P-GRB observed temperature of $ 28.43 \rm{keV}$. We adopted a value for the dyadosphere energy of $E^{e^{\pm}}_{tot}=1.8 \times 10^{53}$ \rm{erg} and a baryon loading of $B= 3.8\times 10^{-3}$. The simulated light curve and spectrum of episode 2 are shown in Figs. \ref{FITTED LIGHT CURVE 2} and \ref{FITTED SPECTRUM}, respectively.

\begin{figure}
\centering
\includegraphics[width=\hsize]{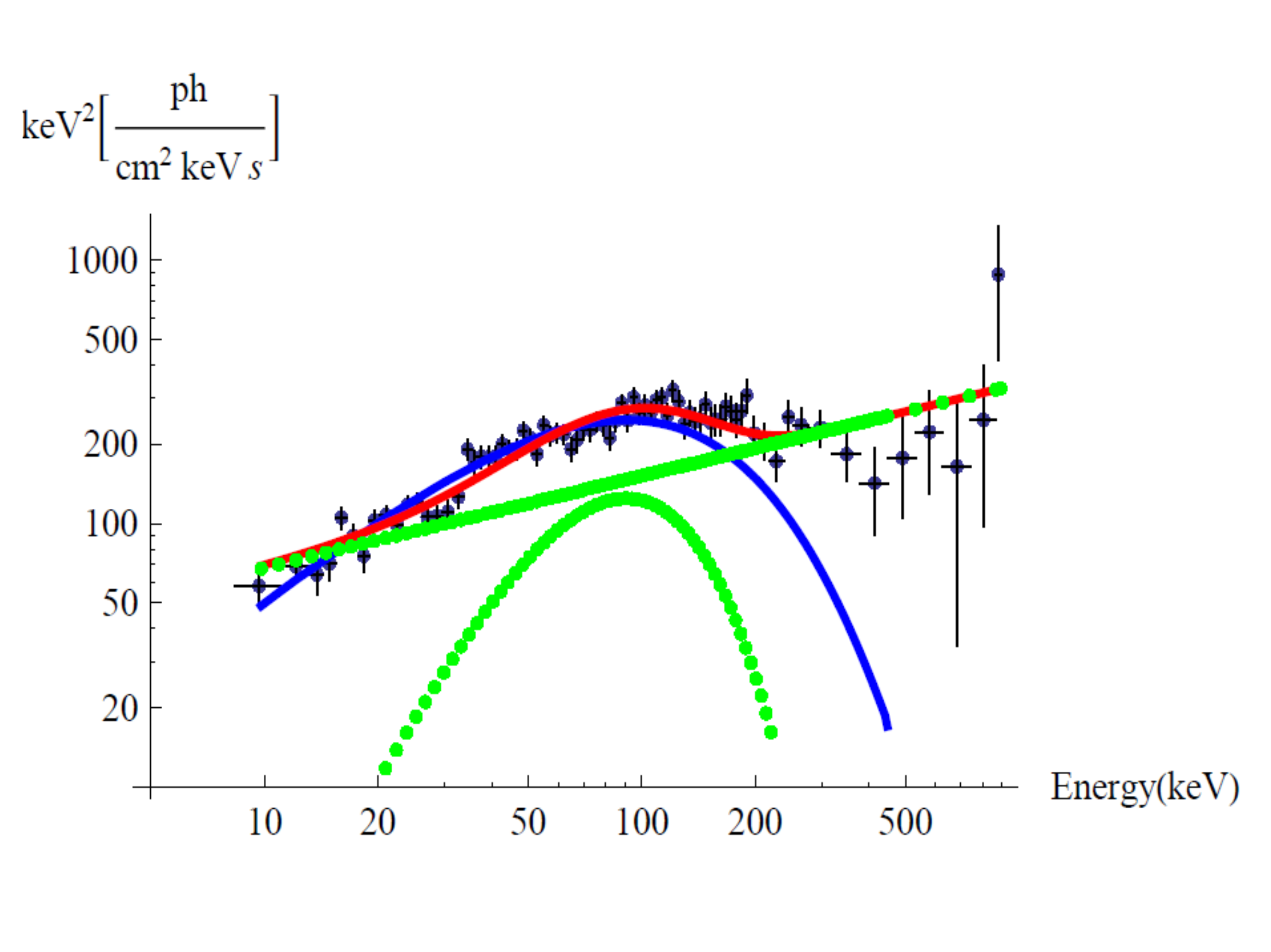}
\caption{Fit of the spectrum of episode 2. The green dotted lines represent the fit of a black body and a power-law components, separately. The red line is the sum of them, calculated with XSPEC (BB+po). The blue line is the fit with the modified black body spectrum given in Eq.(\ref{spettro_barbara}), calculated with the GRBsim numerical code.}
\label{All the fits together}
\end{figure}

Figure \ref{All the fits together} shows the fitted spectrum with different models. We took the data points from the NaI n2 and BGO b0 detectors together. We note there is a good agreement between both fits, in the low and medium energy range. At high energies, the spectrum follows a power-law behavior, which cannot be reproduced by the modified black body model due to the exponential cutoff.

\section{Analysis of the first episode}\label{sec:8}

\begin{figure}
\centering
\includegraphics[width=\hsize]{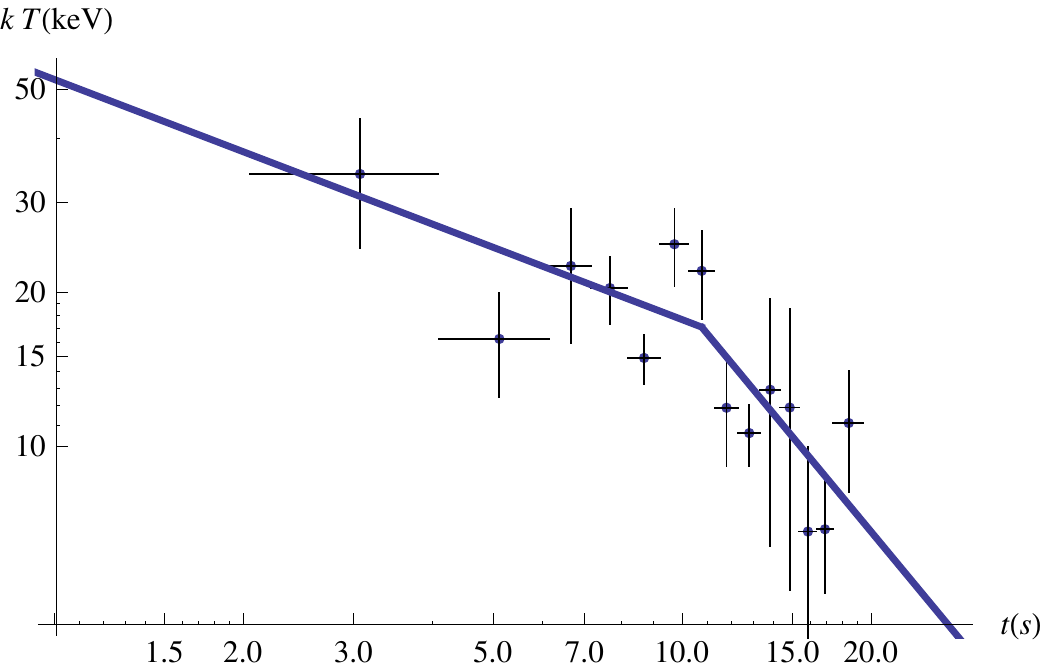}
\caption{Evolution of the observed temperature kT of the BB component. The blue line corresponds to a broken power-law fit. The indices of the first and second power laws are $\alpha=-0.47 \pm 0.34$ and $\beta= -1.48 \pm 1.13$, respectively. The break occurs at 11 s after the trigger time.}
\label{kTepisode1}
\end{figure}

\begin{figure}
\centering
\includegraphics[width=\hsize]{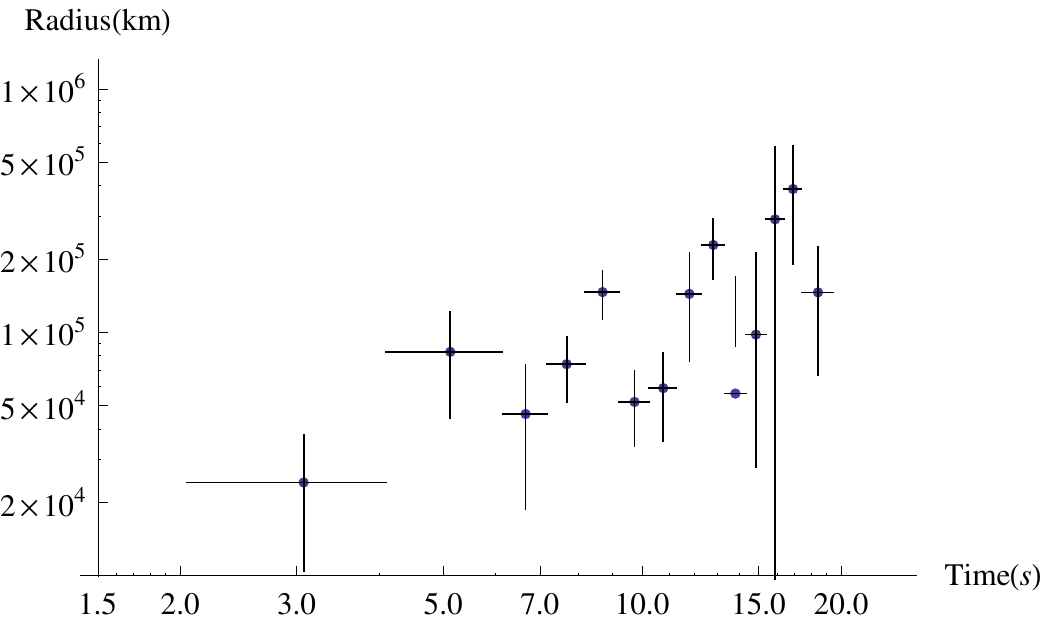}
\caption{Evolution of the radius of the first episode progenitor.}
\label{radius}
\end{figure}

To analyze episode 1 more into detail, in order to identify the nature of this phenomenon, we plotted the temperature of the black body component as a function of time, for the first 20 s of emission (see Fig. \ref{kTepisode1}). We note a strong evolution in the first 20 s of emission which, according to \citet{Ryde2004} can be reproduced by a broken power-law behavior, with $\alpha=-0.47 \pm 0.34$ and $\beta= -1.48 \pm 1.13$ being the indices of the first and second power law, respectively. We also plotted the radius of the most external shell with time (see Fig. \ref{radius}). Following \citet{Luca}, the radius can be written as

\begin{equation}
r_{em}=\frac{\hat{R} D \Gamma}{(1+z)^2},
\end{equation}
where $\hat{\rm{R}}^2= \phi_{obs} /(4 \pi \sigma T_{obs}^4)$ is a parameter, $D$ the luminosity distance, $\Gamma$ the Lorentz factor, and $\phi_{obs}$ the observed flux. We can see that the radius remains almost constant (in fact it increases, but only slightly). From this it is possible to see that the plasma is expanding at nonrelativistic velocities. According to the work of \citet{Arnett}, there is an expansion phase of the boundary layers, while the iron core suffers a contraction. This is due to the presence of strong waves originated while the different shells of the progenitor mix during the collapse phase. This fact confirms the non-GRB nature for the first episode.

\subsection{The X-ray afterglow as a possible redshift estimate ?}

We have seen that GRB 090618 and GRB 101023 share similar properties. They seem to be composed of two different emission episodes, the first being connected to a quasi-thermal process before the collapse of the core, while the second is the canonical GRB \citep[see][]{Ruffini2010a,Ruffini2010AdvSpResSub}.

\begin{figure}      
\centering
\includegraphics[width=\hsize]{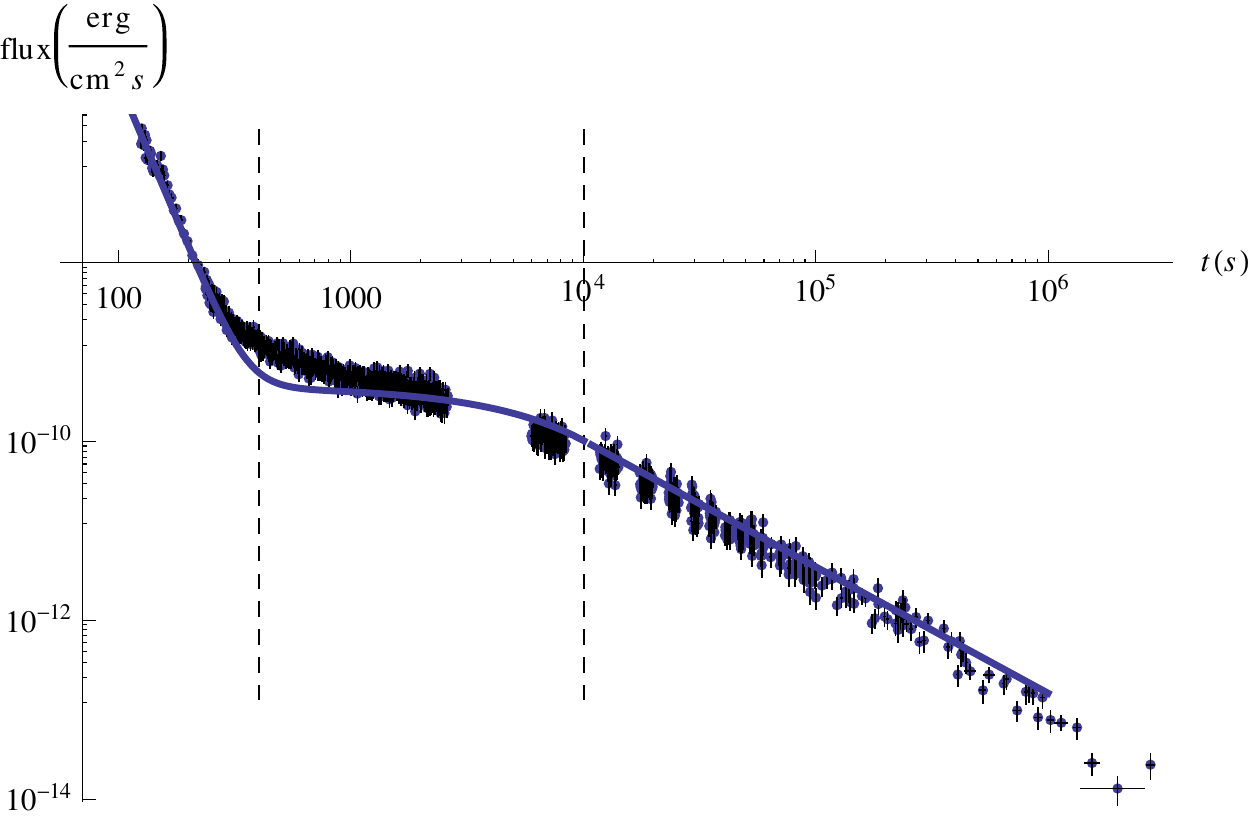}
\includegraphics[width=\hsize]{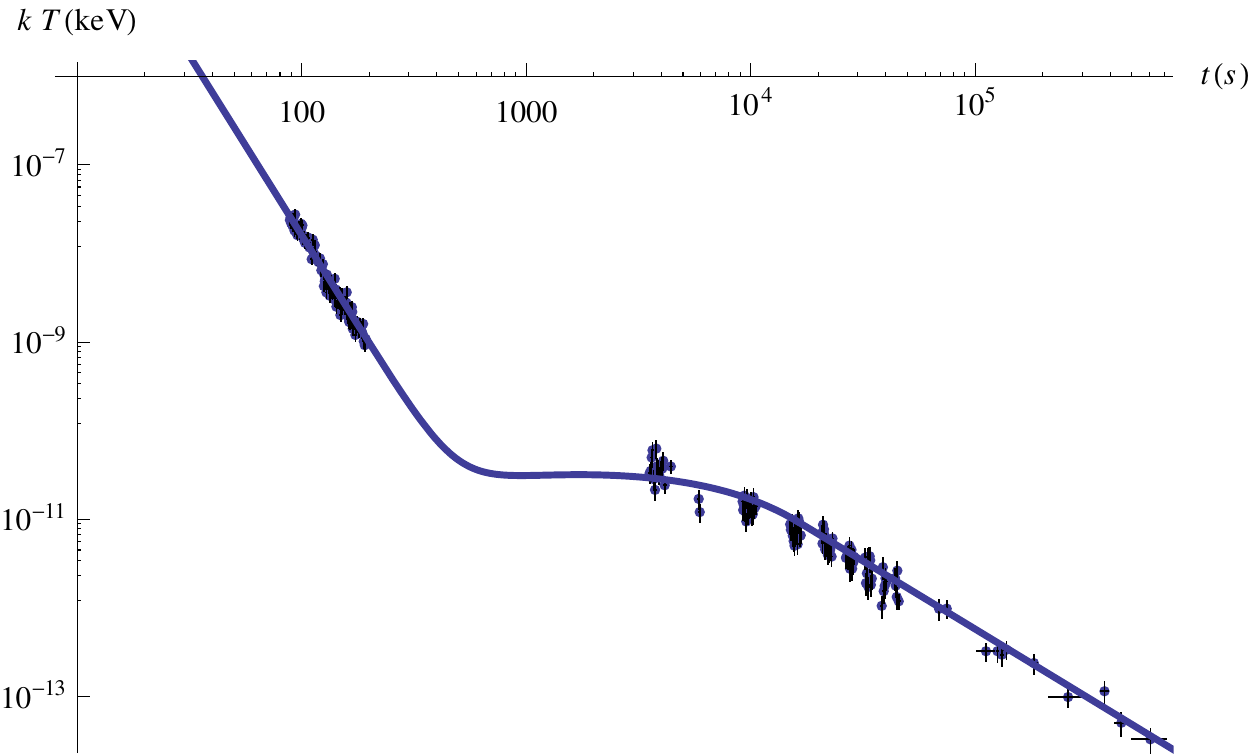}
\caption{The fit of the X-ray afterglow of GRB 090618 (upper panel) and GRB 101023 (lower panel) with the model of \cite{Willingale2007}.}
\label{fig:no27}
\end{figure}

Anyway, if both GRBs were created originated by the same physical mechanism and since the energetics are very similar, considering the value $z = 0.9$ for GRB 101023, we can expect similar luminosity behavior for the X-ray afterglow. Although we have not yet developed a theory for this late afterglow emission, we attempted a simple test that compared the observed X-ray afterglow of both GRBs as if they were located at the same redshift. Since there are different spectral components in the GRB X-ray afterglow, we built the pseudo-redshift light curves for both these different emissions. Thanks to the Swift-XRT observations, we know that the early X-ray afterglow of both GRBs shows a canonical behavior, where the emission can be divided in three distinct parts \citep{Nousek2006}: 1) a first very steep decay, associated with the late prompt emission; 2) a shallower decay, the plateau; 3) a final steeper decay.
At first, we determined for GRB 090618 and GRB 101023 these three time intervals by using the phenomenological function introduced in the work of \citet{Willingale2007}:

\begin{equation}\label{eq:5.1}
 f(t) = \begin{cases}
         F_c exp\left(\alpha_c - \frac{t \alpha_c}{T_c}\right) exp \left(\frac{-t_c}{t}\right), \,\, t < T_c; \\
         F_c \left(\frac{t}{T_c}\right)^{-\alpha_c} exp \left(\frac{-t_c}{t}\right), \,\,\, t > T_c,
        \end{cases}
\end{equation}
which represents the transition from an exponential regime to a power law. This transition occurs at the point $(T_c, F_c)$ where the two functional sections have the same value and gradient. The $\alpha_c$ parameter determines both the time constant of the exponential decay and the temporal decay index of the power law, while the $t_c$ parameter marks the initial rise.
The maximum flux occurs at $t = (t_c T_c/\alpha_c)^{1/2}$. We fit the afterglow data of the two GRBs with this model, and the results of our fits are shown in Fig. \ref{fig:no27}.

\begin{figure}      
\centering
\includegraphics[width=\hsize]{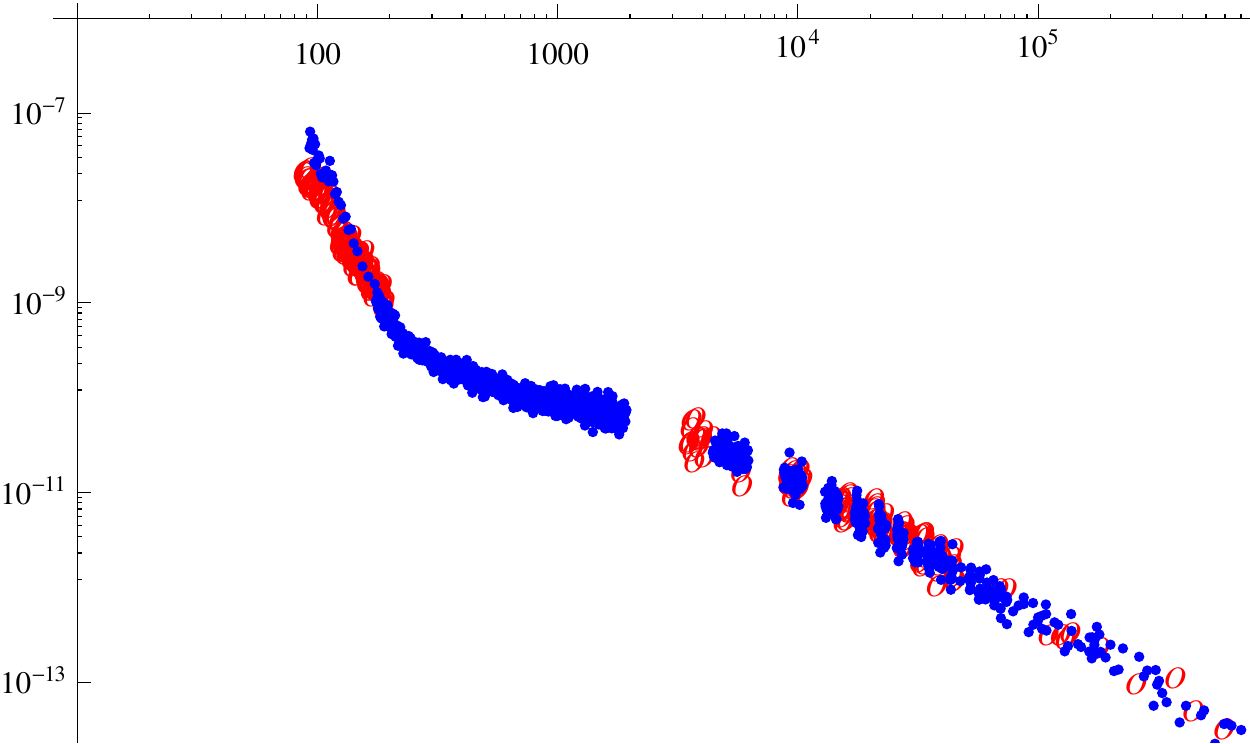}
\caption{The X-ray afterglow of GRB 090618 (blue data) as if it was observed at redshift $z=0.9$ (see text). The X-ray afterglow of GRB 101023 is also shown as comparison (red data). Data on GRB 101023 are missing between $\sim 200$ s and $3550$ s. Where data are present, the superposition is striking.}
\label{fig:no28}
\end{figure}

After the determination of these three time intervals, we built the X-ray light curve of GRB 090618 as if it was observed at redshift z = 0.9, which is our estimate for the redshift of GRB 101023.
The Swift-XRT (which operates in the (0.3 - 10) keV energy range) light curve of GRB 090618 \citep{Evans2007, Evans2009} corresponds to the emission in the rest frame at z = 0.54 in the energy range (0.462 - 15.4) keV, while for GRB 101023 the XRT window corresponds to the range (0.57 - 19) keV.
We must obtain the emission of GRB 090618 in this last energy range, in order to compare the two light curves. At first we made the assumption that the spectrum of each time interval is best fitted by a simple power-law model. This assumption is supported by the hypothesis that the X-ray afterglow comes from a synchrotron emission mechanism \cite{Sarietal1999}, whose spectral emission is represented by a simple power law function. Then, we extrapolated the emission of the afterglow of GRB 090618 in the (0.57 - 19) keV energy range by considering the ratio between the number of photon counts in both energy ranges.
This value corresponds to a conversion factor, which we consider for scaling the intensity of the light curve. We finally amplified, by a term $(1 + z_{101023})/(1 + z_{090618})$, the time interval of emission of GRB 090618, obtaining as a final result the afterglow light curve of GRB 090618 as if it was observed by XRT at redshift 0.9, see Fig. \ref{fig:no28}. It is, most remarkably, a perfect superposition of the light curve emission of both GRBs. This evidence delineates three important aspects:
\begin{itemize}
\item the X-ray afterglow of both GRBs clearly confirms a common physical mechanism for these GRBs;
\item there is ample convergence and redundancy with different methods of determining a value of redshift $z=0.9$ for GRB 101023. There has also been the unexpected result pointing to the late afterglow as a possibly independent redshift estimator;
\item the redshift of GRB 101023 derived by the superposition of the two afterglow curves is consistent with the value of $z = 0.9$, which we have found before.
\end{itemize}

\begin{figure}      
\centering
\includegraphics[width=\hsize]{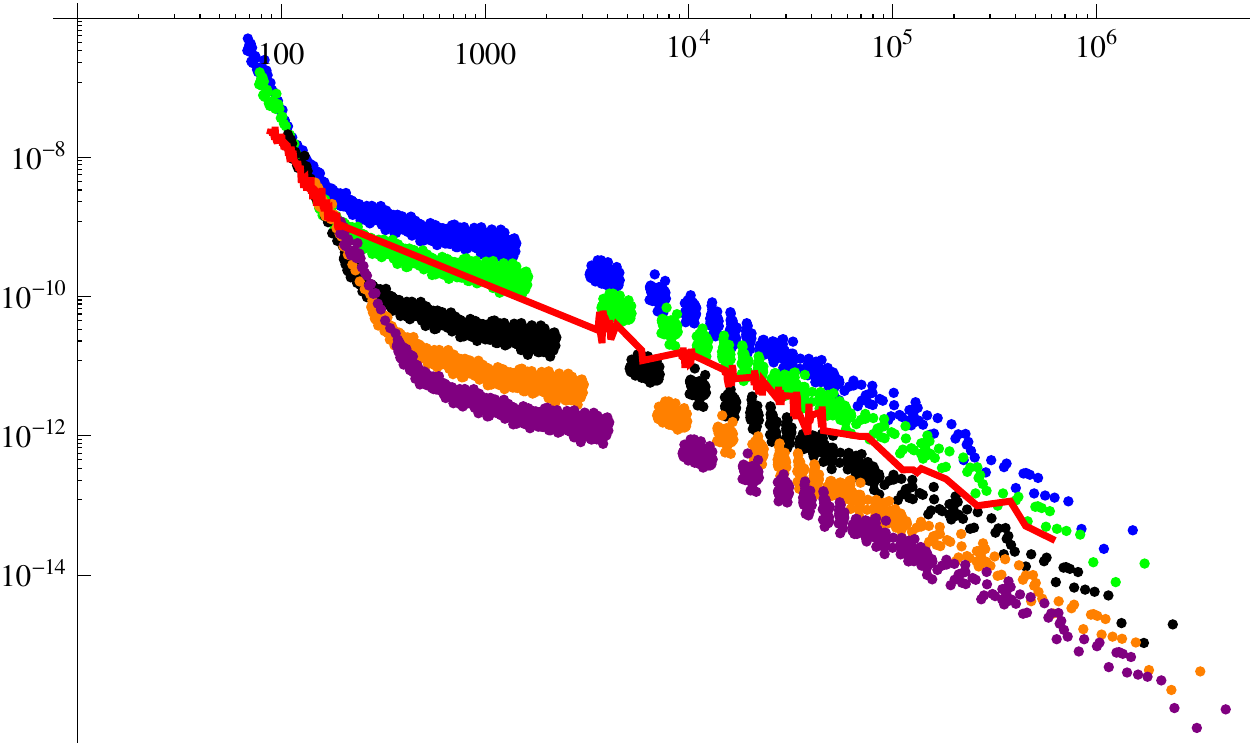}
\caption{The X-ray afterglow of GRB 090618 as if it was observed at different redshifts $z = (0.4, 0.6, 1.2, 2, 3)$, where each color corresponds to a different redshift. The X-ray afterglow of GRB 101023 is also shown for comparison (red data). }
\label{fig:no29}
\end{figure}

This last point led us to do another analysis consisting in the redshift-translation of the X-ray afterglow of GRB 090618 considering different values for the redshift. Following the same procedure and considering five different values for the redshift, $z = (0.4, 0.6, 1.2, 2, 3)$, we see that the X-ray emission of GRB 101023 is compatible with the X-ray afterglow of GRB 090618 as if it bursted between $z = 0.6$ and $z = 1.2$, see Fig. \ref{fig:no29}. Then we conclude that our estimate for the redshift of GRB 101023 of $z = 0.9$ is very reliable.

\section{Conclusions}\label{sec:9}

GRB 101023 is a very interesting source for the following reasons.

1) We find a striking similarity between GRB 101023 and GRB 090618, as can be seen from the light curves. Following the study of GRB 090618, we divided the emission into two episodes: episode 1, which lasts 45 s, presents a smooth emission without spikes that decays slowly with time. Episode 2, of 44 s of duration, presents a spiky structure, composed of a short and faint peak at the beginning, followed by several intense bumps, after which there is a fast decay with time. Episode 2 has all the characteristics of a canonical long GRB.

2) We performed a time-resolved analysis of episode 1. We fitted a black body plus a power-law model and plotted the evolution of the black body component with time. The observed temperature decreases during the first 20 s following a broken power law: the first with index  $\alpha=-0.47 \pm 0.34$ and the second with index $\beta= -1.48 \pm 1.13$, see Sec. 9. This behavior is very similar to GRB 090618.

3) In the absence of a direct measurement of the redshift to the source, we have inferred it from several empirical methods. First, following the work of \citet{Grupe}, which considers the hydrogen equivalent column density in the direction of the source, we obtained an upper limit of $z<3.8$. Then we performed a spectral analysis to episode 2, fitting a Band model. From the peak energy $E_{peak}$ and using the Amati relation under the hypothesis that episode 2 is a canonical long GRB, we constrain the value of the redshift to be between 0.3 and 1.0. Finally, using the parameters of the Band model and following the work of \citet{Atteia}, we determine a value of the redshift of $z=0.9 \pm 0.084(stat.) \pm 0.2 (sys.)$. The three methods are consistent, so we assumed for the redshift of this source  $z=0.9$. 

4) From the knowledge of the redshift of the source, we have analyzed episode 2 within the fireshell model. We determined a total energy $E_{iso}=1.79 \times 10^{53}$ erg and a P-GRB energy of $2.51 \times 10^{51}$ erg, which we used to simulate the light curve and spectrum with the numerical code GRBsim. We find a baryon load $B=3.8 \times 10^{-3}$ and, at the transparency point, a value of the laboratory radius of $1.34 \times 10^{14}$ cm, a theoretically predicted temperature of $kT_{th}=13.26$ keV (after cosmological correction) and a Lorentz gamma factor of $\Gamma=260.48$, confirming that episode 2 is indeed a canonical GRB.

5) From the knowledge of the redshift, we can also evaluate the flux emitted by episode 1, and from the observed black body temperature, infer the radius of the black body emitter and its variation with time, see Fig. 12. We saw that it increases during the first 20 s of emission, with a velocity $\sim 1.5 \times 10^4$ km/s. In analogy with GRB 090618, we concluded that episode 1 originates in the last phases of gravitational collapse of a stellar core, just prior to the collapse to a black hole. We call this core a ``proto-black hole'' \citep{Ruffini2010a}. Immediately afterwards, the collapse occurs and the GRB is emitted (episode 2). 

6) Finally, we performed the following test. Owing to the similarities between GRB 101023 and GRB 090618 regarding morphology and energetics, we expect them to be created by the same physical mechanism, so we compared the late observed X-ray afterglow of both GRBs as if they were located at the same redshift; i.e, we built the light curve of GRB 090618 (of $z=0.54$) as if it had redshift $z=0.9$, extrapolating it to the XRT energy window of GRB 101023. We found a surprising perfect superposition of the light curves for z=0.9, receiving a further confirmation of the correctness of the cosmological redshift determination. The same procedure for the redshift determination will be repeated for sources with a spectroscopical-determined redshift, as a further check of our proposal. This result points to a possible use of the late afterglow as a distance indicator. 

We concluded that GRB 101023 and GRB 090618 have striking analogies and are members of a specific new family of GRBs developing out of a single core collapse. It is also appropriate to remark that this new kind of source does not present any GeV emission. The existence of precise scaling laws between these two sources opens a new window on the use of GRBs as distance indicators. We will go on to identify additional sources belonging to this family. This new paradigm is also being applied to sources at very high redshift to see how the absence of a signal under the threshold can affect the theoretical interpretation. We are also considering the possibility that proto-neutron stars in addition to proto-black holes may exist in the case of supernovae or hypernovae. Particularly interesting in this respect is the work of \citet{Soderberg2008} showing the X-ray emission prior to SN events, which may relate the observed X-ray emission prior to SN 2008D to episode 1 in GRB 090618 and GRB 101023. In this sense we are revisiting our considerations of GRB 980425 \citep[see e.g.][]{Fraschetti2004,Fraschetti2005,Ruffini2004,Ruffini2007,Bernardini2008}, as well as of GRB 030329 \citep{2004AIPC..727..312B,2005tmgm.meet.2459B} and GRB 031203
\citep{2005ApJ...634L..29B,Ruffini2007,2008ralc.conf..399R}.

\acknowledgements

We thank the Swift and Fermi teams for their support. This work made use of data supplied by the UK Swift Data Centre at the University of Leicester. We are especially grateful to an anonymous referee for her/his important remarks which have improved the presentation of our results.

\end{document}